\documentclass[pdflatex,sn-mathphys-num]{sn-jnl}

\usepackage{graphicx}%
\usepackage{multirow}%
\usepackage{amsmath,amssymb,amsfonts}%
\usepackage{amsthm}%
\usepackage{mathrsfs}%
\usepackage[title]{appendix}%
\usepackage{xcolor}%
\usepackage{textcomp}%
\usepackage{manyfoot}%
\usepackage{booktabs}%
\usepackage{algpseudocode}%
\usepackage{listings}%

\usepackage{amsthm}
\usepackage{comment}

\usepackage[ruled, linesnumbered]{algorithm2e}

\usepackage{bm}

\usepackage{subfig}

\usepackage{varioref}

\usepackage{courier}
\usepackage{color}
\usepackage{colortbl}
\usepackage{multirow,hhline}
\usepackage{booktabs}
\usepackage[figuresright]{rotating}

\usepackage{makecell,rotating}
\usepackage{array}
\usepackage{extarrows}
\usepackage{bbding}
\usepackage{subfig}

\usepackage{amssymb}

\usepackage{enumerate}

\usepackage{wrapfig}

\usepackage{tikz}
\newcommand*\circled[1]{\tikz[baseline=(char.base)]{
            \node[shape=circle,draw,inner sep=0.8pt] (char) {#1};}}

\definecolor{Gainsboro}{RGB}{220,220,220}
\definecolor{FloralWhite}{RGB}{255,250,240}
\definecolor{Honeydew}{RGB}{240,255,240}
\definecolor{Lavender}{RGB}{230,230,250}

\theoremstyle{thmstyleone}%
\newtheorem{theorem}{Theorem}
\newtheorem{lemma}{Lemma} %

\theoremstyle{thmstyletwo}%
\newtheorem{remark}{Remark}%

\theoremstyle{thmstylethree}%
\newtheorem{definition}{Definition}%
\newtheorem{problem}{Problem}

\begin{document}

\title[Article Title]{Game Theory Based Community-Aware Opinion Dynamics}


\author[1]{\fnm{Shanfan} \sur{Zhang}}\email{zhangsfxajtu@gmail.com}

\author[2]{\fnm{Xiaoting} \sur{Shen}}\email{njdnsxt@126.com}

\author*[3]{\fnm{Zhan} \sur{Bu}}\email{zhanbu@nau.edu.cn}

\affil[1]{\orgdiv{School of software Engineering}, \orgname{Xi’an Jiaotong University}, \orgaddress{\city{Xi’an}, \postcode{710100}, \state{Shaanxi}, \country{China}}}

\affil[2]{\orgdiv{School of Automation}, \orgname{Nanjing University of Information Science \& Technology}, \orgaddress{\city{Nanjing}, \postcode{210040}, \state{Jiangsu}, \country{China}}}

\affil[3]{\orgdiv{School of Computer Science}, \orgname{Nanjing Audit University}, \orgaddress{\city{Nanjing}, \postcode{211815}, \state{Jiangsu}, \country{China}}}



\abstract{Individuals consistently encounter and interact with diverse opinions on a wide range of topics, spanning from public politics to everyday experiences. Examining the mechanisms underlying the formation and evolution of opinions within real-world social systems, which consist of numerous individuals, can provide valuable insights for effective social functioning and informed business decision-making. The focus of our study is on the dynamics of opinions inside a networked multi-agent system. We provide a novel approach called the Game Theory Based Community-Aware Opinion Formation Process (\emph{GCAOFP}) to accurately represent the co-evolutionary dynamics of communities and opinions in real-world social systems. The \emph{GCAOFP} algorithm comprises two distinct steps in each iteration. 1) The \textbf{\emph{Community Dynamics Process}} conceptualizes the process of community formation as a non-cooperative game involving a finite number of agents. Each individual agent aims to maximize their own utility by adopting a response that leads to the most favorable update of the community label. 2) The \textbf{\emph{Opinion Formation Process}} involves the updating of an individual agent's opinion within a community-aware framework that incorporates bounded confidence. This process takes into account the updated matrix of community members and ensures that an agent's opinion aligns with the opinions of others within their community, within certain defined limits. The present study provides a theoretical proof that under any initial conditions, the aforementioned co-evolutionary dynamics process will ultimately reach an equilibrium state. In this state, both the opinion vector and community member matrix will stabilize after a finite number of iterations. In contrast to conventional opinion dynamics models, the guaranteed convergence of agent opinion within the same community ensures that the convergence of opinions takes place exclusively inside a given community. This unique characteristic allows for the creation of diverse opinions within the network while also ensuring that each agent's opinion ultimately converges to a local consensus. A number of detailed tests were done on real-world networks to validate the effectiveness and scalability of \emph{GCAOFP}. Subsequently, a comparison was made between \emph{GCAOFP} and state-of-the-art graph clustering methods in the context of community detection.}




\maketitle

\section{Introduction}\label{sec1}



The study of opinion dynamics in social networks is a significant and well-established interdisciplinary field of research. It has garnered considerable attention and a growing body of literature \cite{literature-1, literature-2, literature-3, literature-4}, particularly in the development of agent-based models \cite{agent-based-1, agent-based-2, agent-based-3, agent-based-4}. These models are utilized to investigate the underlying mechanisms that contribute to the building of consensus and clustering of opinions within groups. The opinions of a group of agents, who interact with each other through a network structure, will undergo a process of fusion based on predetermined rules until all members of the group reach a state of equilibrium \cite{opinion-survey-1}. This process results in the emergence of collective opinions that can be categorized as either consensus, polarization, or fragmentation \cite{opinion-dynamic}.

Opinion dynamics has traditionally been investigated within the framework of networked multi-agent systems \cite{multi-agent}. In this context, individuals, represented as agents, engage in active exchange of opinions with their neighboring agents while concurrently receiving opinions from them \cite{change_opinion}. Agents regularly revise their opinions in order to adapt to changes in their surrounding environment. In earlier studies, it is usual to hypothesis about consensus issues, where a collective of agents aim to attain a shared objective \cite{assum-consensus-1, assum-consensus-2}. However, in certain scenarios like political elections, agents do not desire to achieve consensus and are unlikely to converge towards a unified outcome of the underlying process. Motivated by the aforementioned facts, some studies are dedicated to examining the inclusion of disagreement with consensus within the framework of opinion dynamics \cite{disagree-1, disagree-2}. One such model that has garnered significant attention from scholars and produced seminal outcomes is the Hegselmann-Krause model (\textbf{HK}) \cite{HK}. For instance, increasing the confidence level will result in a reduced number of opinion clusters. In the extreme case when the confidence level value is sufficiently strong (e.g., 1), there will be a single cluster among the agents, indicating that all agents will ultimately reach a consensus \cite{HK-consensus}.

In our perspective, a deficiency observed in classical opinion dynamics models pertains to the inadequate and inefficient incorporation of social networks. Put otherwise, their assumption is limited to the exchange of opinions solely among networked neighbors, neglecting to account for higher-order connections between agents, such as communal bonds \cite{higher-order}. Researchers have shown increasing interest in the substantial influence of multiple connections between agents on the dynamics of opinions in recent years \cite{multiple-connections-1, multiple-connections-2}. Aris et al. \cite{structure-example-1} propose a modeling approach to examine online social networks and investigate the influence of network structure on opinion polarization. Henrique et al. \cite{structure-example-2} present a mathematically feasible method to analyze the interaction between opinion dynamics and the underlying network structure. They also analyze the duration required for the system to reach a consensus on the dominant opinion.

The theory in \cite{structure-opinion} suggests that the coexistence of opinions within a social network often presents the locality effect, in which an opinion or a fad is limited to specific groups and can not infect the whole society. This implies that community structure could have an impact on opinion dynamics, as members in real-world social systems usually share common interests with interactive users within the same community, thus they tend to trust neighbors within the same community more than neighbors outside of their community \cite{community-trust}. In this paper, we introduce a novel and powerful \textbf{\emph{G}}ame Theory Based \textbf{\emph{C}}ommunity-\textbf{\emph{A}}ware \textbf{\emph{O}}pinion \textbf{\emph{F}}ormation \textbf{\emph{P}}rocess (in short as \emph{GCAOFP} henceforth) to study the co-evolving dynamics of communities and opinions in social networks. Although some researches have considered applying opinion dynamics to solve the community detection problems \cite{community-given-1, community-given-2}, they ignored the fact that, in the absence of predefined communities, the evolution of opinions and the formation of communities dynamically interact and evolve in many practical situations, until a stable state is ultimately reached. \emph{GCAOFP} believes that agents with similar opinions are more likely to belong to the same community, and agents trust the opinions of neighbors in the same community more. Therefore, every time agents update their opinions, they should consider the impact of the current communities, and the agent's opinion is also considered when dividing the community label for the agent. \emph{GCAOFP} also utilizes the concept of bounded confidence, that is, agents will only communicate and accept the opinions of their neighbors within their confidence ranges. The main contributions of this study can be summarized as follows:

\begin{enumerate}[1)]

\item In this study, we examined the process of co-evolution between community partitioning and opinion evolution inside social networks. In order to replicate the phenomenon of inter-community discrimination observed in real-world social systems, we have proposed a theoretical construct known as "community susceptibility." This construct serves to quantify the degree to which individuals place faith in the judgments of their neighbors who belong to the same community.

\item The dynamics of community development were conceptualized as a non-cooperative game with a finite number of agents who strive to enhance social welfare. The opinion dynamics are implemented by a mechanism similar to the \textbf{HK} model, wherein each agent updates their opinion by calculating a weighted average of their own internal opinion and the sympathetic opinions of their neighboring agents.

\item A scalable algorithm with linear time complexity has been created for the co-evolving dynamics of communities and opinions. Theoretical evaluations indicate that the method under consideration has the capability to reach a state of convergence, whereby both the opinion vector and the community member matrix exhibit stability, within a finite number of iterations.

\item The efficacy, scalability, and parameter sensitivity of \emph{GCAOFP} were proved through the application of quantitative analysis. Numerous empirical investigations were undertaken to explore the dynamics of opinion evolution and community detection in various real-world social networks. The findings indicate that the \emph{GCAOFP} model exhibits a significant level of competitiveness when compared to established benchmark models.

\end{enumerate}

\section{Symbol description}

In this article, we use the bold and italic uppercase letters are used to denote matrices (i.e., $\mathbf{W}$) and sets (i.e., $\mathcal{N}$) respectively. The bold lowercase letters to denote vectors (i.e., $\mathbf{x}$). In particular, we denote by $\mathbf{W}_{i}$ the $i$-th row of $\mathbf{W}$, $\mathbf{W}_{ij}$ the $\left( i, j \right)$-th element of matrix $\mathbf{W}$, $\mathbf{x}_{i}$ the $i$-th element of $\mathbf{x}$. The \emph{Hadamard product} between matrices $\mathbf{D} = \left [ d \right ]_{n\times n}$ and $\mathbf{W}= \left [ w \right ]_{n\times n}$ is denoted as $\mathbf{D} \odot \mathbf{W}$, and the \emph{matmul product} between matrices $\mathbf{A} = \left [ a \right ]_{m\times n} $ and $\mathbf{\mathbf{B}} = \left [ b \right ]_{n\times k}$ is denoted as $\mathbf{A} \mathbf{\mathbf{B}} = \left [ c \right ]_{m \times k}$. In addition, we denote $tr(\mathbf{W})$ as the trace of $\mathbf{W}$, $\mathbf{W}^{\mathsf{T}}$ as the transpose of $\mathbf{W}$, $\mathbf{1}_{n}$ as the $n \times n$ identity matrix, $1_{n}$ as an all-one column vector with $n$ elements, and $\mathbf{1}^{n} = 1_{n} 1_{n}^{\mathsf{T}}$ as the all-one $n \times n$ matrix.

Opinion dynamics studies the evolution of opinions in a network through the interaction between agents. Consider a social system with a fixed set of agents $\mathcal{N} = \left \{ 1,2,\dots ,n \right \}$, the relationship between agents is unidirectional, so the relationships among them can be represented by an $n \times n$ asymmetric matrix $\mathbf{W}$, where each non-diagonal element $\mathbf{W}_{ij}$ measures the confident level of agent $i$ in the opinion of agent $j$. $\mathbf{W}$ must also meet the following characteristics: (i) without self-loops, i.e., $\forall i \in \mathcal{N}, \mathbf{W}_{ii} = 0$; and (ii) irreducible, i.e., all nodes in the associated directed network are strongly connected to each other. Given a topic, at each time $T \in \mathbb{R} ^{+}$, agent $i \in \mathcal{N}$ can hold a opinion $\mathbf{x}_{i}\left ( T \right ) \in \left [ 0, 1 \right ] $, and the opinions of all members in the society can be represented by a opinion vector $\mathbf{x} \left ( T \right ) \in \left [ 0, 1 \right ]^{n}$.

We also define a community member matrix $\mathbf{s} \left ( T \right ) \in \left \{ 0,1 \right \}^{n\times K}$ to indicate the division of agents at time $T$, where $\mathbf{s}_{ik}\left ( T \right ) = 1$ implies that agent $i$ belongs to the $k$-th community, and 0 otherwise. Suppose that every agent can only join in one community, then we have $\forall i\in N, {\textstyle \sum_{k=1}^{K}}\mathbf{s}_{ik}\left ( T \right ) =1$. We denote the $i$-th row of $\mathbf{s}\left ( T \right )$, i.e., $\mathbf{s}_{i}\left ( T \right )$, as the community membership vector of agent $i$; thus, the quantity $\mathbf{s}_{i}\left ( T \right )\mathbf{s}_{j}^{\mathsf{T}}\left ( T \right ) = 1$ if agents $i$ and $j$ are in the same community, and 0 otherwise. Finally, we define matrix $\Lambda = diag\left ( \lambda_{1},\lambda_{2}, \dots ,\lambda_{n}\right )$ to represent the relative trust level of the agents to their neighbors within the same community, i.e., any entry $\lambda_{i} \in \left ( 1,+  \infty  \right )$ indicating how much agent $i$ trusts the intra-community neighbors (for inter-community neighbors, agents set their trust to 1 by default).

\section{Related Work}

\subsection{The Hegselmann-Krause model}

Let $\varrho \in \left ( 0, 1 \right )$ be the range of confidence, the \textbf{HK} model selects "trusted" neighbors for any agent $i \in \mathcal{N}$ at every time point $T \in \mathbb{N}$ according to the opinion vector $\mathbf{x}\left ( T \right )$: $\mathcal{I}_{i}=\left \{ j: \left | \mathbf{x}_{j}\left ( T \right )-\mathbf{x}_{i}\left ( T \right ) \right | < \varrho \wedge \mathbf{W}_{ij}> 0 \right \} \cup \left \{ i \right \} $. Each agent accepts the average opinion of all trusted neighbors as her own opinion for the next moment, i.e., 
\begin{equation}
	\begin{aligned}
		\label{HK_DEFINE}
			\mathbf{x}_{i}\left ( T+1 \right ) = \frac{1}{\left | \mathcal{I}_{i}  \right | }  \sum_{j \in \mathcal{I}_{i}} \mathbf{x}_{i}\left ( T \right )
	\end{aligned}
\end{equation}

The \textbf{HK} model is a typical bounded confidence model in which the assumption of influence weights depending on the evolving opinion distance. Arnab et al.\cite{HK-convergence} prove that when agent's opinion is one-dimensional, the convergence time of the \textbf{HK} model in one dimension is at least $\mathcal{O} \left (  n \right )$ and at most $\mathcal{O} \left (  n^{3} \right )$.

\subsection{Community Detection based on Potential Game}

Community detection refers to dividing agents into multiple communities according to the connection pattern in the network, so that agents within the community are tightly connected while there are only a few connections between communities. Mainstream community detection algorithms either set global optimization goals such as \emph{modularity} \cite{MODULARITY} and \emph{betweenness} \cite{betweenness} for detection, or only look for sub-components with a specific predetermined structure, such as \emph{k-clique} \cite{k-clique}. In real social networks, however, communities are formed organically from bottom up, without a centralized authority leading the implementation of a global objective.

The community detection algorithms in the context of non-cooperative game theory modeled the interactions between agents as a potential game in which agents are assumed to be selfish agents committed to maximizing their own utility. The most competitive model along this line we think is \emph{GLEAM} \cite{GLEAM} which is capable of handling large networks with hundreds of millions of agents. \emph{GLEAM} defines a utility function associated with any agent $i$ as $u_{i}\left ( \mathbf{s}\left ( T \right ) \right ) = \frac{1}{M} \sum_{j}\mathbf{\mathbf{B}}_{ij}\mathbf{s}_{i}\left ( T \right ) \mathbf{s}_{j}^{\mathsf{T}}\left ( T \right )$, where $\mathbf{B}$ is the $n \times n$ matrix with elements $\mathbf{\mathbf{B}}_{ij}= \mathbf{A}_{ij}-\frac{d_{i}d_{j}}{2M}$, $\mathbf{A} \in \left \{ 0,1 \right \} ^{n\times n}$ is the adjacency matrix of an undirected network, $d_{i} = \sum_{j} \mathbf{A}_{ij}$ stands as the degree of agent $i$, and $M=\sum_{ij} \mathbf{A}_{ij}$ is the total number of connections. \emph{GLEAM} proves that non-cooperative game under this definition can well match the potential game with potential function $\phi \left ( \mathbf{s}\left ( T \right ) \right ) = \frac{1}{2M}tr\left ( \mathbf{s}^{\mathsf{T}}\left ( T \right ) \mathbf{B} \mathbf{s}\left ( T \right ) \right )$.

\subsection{Community-related opinion dynamics models}

From the microscopic perspective of the evolutionary process, opinion evolution and community division can influence each other. Two recent proposed co-dynamics models on how community partitioning in networks affects the evolution of group opinions deserve our attention. \emph{GK-Means} \cite{GK-MEANS} first identifies those opinion leaders with high local reputation, then employs a dynamics game model to find the locally Pareto-optimal community structure, and finally finds the set of trusted neighbors for each agent according to the current opinion vector, and updates the agents' opinion adopting a \emph{DF}\cite{DF-MODEL}-like dynamical process. \emph{GK-Means} repeats the above three-coupled phases until the opinion vector converges to a relatively stable state. \emph{LPA-HK} \cite{LPA-HK} integrates the classical \textbf{L}abel \textbf{P}ropagation \textbf{A}lgorithm (\emph{LPA} \cite{LPA}) and standard \emph{HK} models to establish a bounded confidence model with a community detection mechanism. Agents will only accept neighbors' opinions and community labels when their opinion difference within the confidence level. In addition, during the opinion update phase, \emph{LPA-HK} also defines a hyperparameter $\mu \in \left ( 0, 1 \right )$ to account for the impact of the current communities.

\section{Methodology}

Within the framework of the social system, agents are consistently exposed to perspectives from their immediate neighbors and subsequently modify their own opinions in response. Research findings indicate that there exists a positive association between agents sharing similar opinions and their membership in the same community. Moreover, agents tend to exhibit a greater inclination to trust the opinions of their neighbors who belong to the same community. Additionally, it has been observed that agents are inclined to accept the opinions of their neighbors only if they fall within their confidence ranges, while selectively disregarding the opinions of other neighbors. The phenomenon of "information cocoons" \cite{information-Cocoons} frequently arises in the real world, as observed in the field of psychology. Motivated by these facts, the present study employed game theory to investigate the optimal community partitioning that maximizes social welfare. Additionally, a novel mechanism known as the \emph{GCAOFP} model was developed, which incorporates community awareness and bounded confidence opinion dynamics. The model incorporates various parameters, including the degree of similarity in opinions, the threshold of bounded confidence, and the level of trust among agents. Every individual possesses not just a personal viewpoint but also an associated categorization that symbolizes their affiliation with a particular social group. The phenomenon of opinion development is closely intertwined with the proliferation of labels within the realm of social networks, and these two factors exert a reciprocal influence on each other.

The \emph{GCAOFP} process consists of two distinct stages, namely the \textbf{Community Dynamics Process} and the \textbf{Opinion Formation Process}. This research presents an analysis of both the community labels and the opinions given by the agents. In the context of social interactions, it is well observed that individuals tend to initially assign distinct sensitivity coefficients to their neighbors based on their community labels, then afterwards consider their perspectives. This empirical belief is widely acknowledged. During the stage of \textbf{Community Dynamics Process}, all agents within the social network are chosen and their label update is executed iteratively until no further improvement in utility can be achieved by altering their community labels. The synchronization of the agents' update is performed. In the stage of \textbf{Opinion Formation Process}, it is postulated that all agents possess an identical level of confidence, which diminishes with each iteration. Furthermore, each agent exclusively takes into account the opinions of neighboring agents falling within the bounds of their confidence level. The update rule ensures that the opinion of each agent will converge to a stable value within a finite number of iterations. The step-by-step instructions for the proposed community-aware opinion dynamics are outlined in Algorithm \ref{alg}.

\begin{algorithm}[htbp]
\caption{\emph{Game Theory Based Community-Aware Opinion Formation Process}}
\label{alg}
\KwIn{ Influence matrix $ \mathbf{W} $, diagonal matrix $\Lambda$, initial opinion vector $\mathbf{x} \left ( 0 \right )$, equilibrium initial community membership matrix $\mathbf{S} \left ( 0 \right )$, tuning parameters $\beta$ and $\gamma$.}

\KwOut{Equilibrium opinion vector $\mathbf{x}^{\ast}$ and equilibrium community membership matrix $\mathbf{S}^{\ast}$.}

$t \gets 0$ \;
$\mathbf{S}^{\ast} \gets  \mathbf{S} \left ( 0 \right ) $ \;

\While{True}{ 
    $\mathbf{E} \left ( t \right ) \gets \left | \mathbf{x}1_{n}^{\mathsf{T}} - 1_{n}\mathbf{x}^{\mathsf{T}} \right |$ \;
    $\mathbf{D} \gets \Psi - \mathbf{E} $ \;
    $\mathbf{B} \gets ReLu\left(\mathbf{D}  \right )$ \;

    
    $\mathbf{Y} \left ( t \right ) \gets \left ( \Lambda \mathbf{D} - \mathbf{B} \right ) \odot \mathbf{W} + \left ( \mathbf{D}^{\mathsf{T}} \Lambda - \mathbf{B}^{\mathsf{T}} \right ) \odot \mathbf{W}^{\mathsf{T}} $\;

    $\mathbf{S} \gets \mathbf{S}^{\ast} $ \;
    
    $\phi ^{\ast }  \left ( \mathbf{S} \right ) = \frac{1}{2} tr(\mathbf{S}^{\mathsf{T}} \mathbf{Y} \left ( t \right )\mathbf{S})$  \;

    $\Delta \phi ^{ \ast} \gets  \infty $   \;

    \While{$\Delta \phi ^{ \ast} \ne 0 $}{
        $\Delta \phi ^{ \ast} \gets 0 $  \;
        \ForEach{ $ i \in \mathcal{N}$}{%
            $\mathcal{F}_{i} \gets \left \{ e_{\left ( k \right )}^{\mathsf{T}}: \mathbf{W}_{ij} + \mathbf{W}_{ji}> 0, \mathbf{S}_{j}= e_{\left ( k \right )}^{\mathsf{T}}\right \} \cup \left \{ \mathbf{S}_{i} \right \} $  \;

            $\mathbf{S}_{i}^{\diamond} \gets argmax_{e_{\left ( k \right )}^{\mathsf{T}} \in \mathcal{F}_{i}}\textstyle \sum_{j \ne i} e_{\left ( k \right )}^{\mathsf{T}}\mathbf{S}_{j}^{\mathsf{T}}\mathbf{Y}_{ij} \left ( t \right )$ \;

            $\Delta \phi ^{ \ast} \gets \Delta \phi ^{ \ast} + \frac{1}{2}\sum_{j \ne i} \left ( \mathbf{S}_{i}^{\diamond} - \mathbf{S}_{i} \right ) \mathbf{S}_{j}^{\mathsf{T}}\mathbf{Y}_{ij} \left ( t \right ) $   \;

            $\mathbf{S}_{i}  \gets \mathbf{S}_{i}^{\diamond} $  \;
            

            
        }
        \tcc{one can output $\phi ^{\ast } \left ( \mathbf{S} \right )$ to visualize the change of potential value during the update. Theoretically, $\phi ^{\ast } \left ( \mathbf{S} \right )$ will be incremental.}
        $\phi ^{\ast }  \left ( \mathbf{S} \right ) = \phi ^{\ast }  \left ( \mathbf{S} \right ) + \Delta \phi ^{ \ast} $ \;
    } 
    
    $\Xi ^{T} \gets \beta \gamma ^{T} \mathbf{1}^{n} $  \;
    

    $\Theta \gets \left [ \mathbf{1}^{n} +\left ( \Lambda - \mathbf{1}_{n} \right ) \mathbf{S} \mathbf{S}^{\mathsf{T}} \right ]  \odot \mathbf{W} \odot Relu \left ( \Xi ^{T} - \mathbf{E} \right )$ \;
    
    $\Phi \gets \Theta \oslash \left [ \Theta \mathbf{1}^{n} + \mathbf{1}^{n} \right ] $ \;

    $\Phi^{\prime} \gets 1_{n} \oslash \left [ \Theta 1_{n} + 1_{n}\right ] $ \;

    $\mathbf{x}^{\ast} \gets \Phi^{\prime} \odot \mathbf{x}  + \Phi \mathbf{x}  $\;
    
    $\mathbf{S}^{\ast} \gets \mathbf{S} $ \;

    \If{$ min \left | \mathbf{x}^{\ast} - \mathbf{x} \right | < \epsilon $ }{
      break \;
    }
    $\mathbf{x} \gets \mathbf{x}^{\ast}$  \;
    $t \gets t+1$  \;
}

\textbf{Return} $\mathbf{x}^{\ast}$ and $\mathbf{S}^{\ast}$ \;

\end{algorithm}

\subsection{Community Dynamics Process}

At any time $T \in \mathbb{R} ^{+}$, we have the opinion vector $\mathbf{x} \left ( T \right )$ and the community member matrix $\mathbf{S}\left( T \right)$, and in the community update stage, our goal is to find the most reasonable community structure so that the opinions between agents within the same community are as similar as possible. We formulate this community partition problem as as a non-cooperative game played by rational agents, where the goal of each agent is to adjust her community label to belong to the same community as many neighbors with similar opinions as possible.

\begin{definition}\label{def:CDP}[\textbf{Community Dynamics Process (CDP)}] Given the agents set $\mathcal{N}$, opinion vector $\mathbf{x} \left ( T \right )$, influence matrix $\mathbf{W}$ and relative trust matrix $\Lambda$, the proposed \textbf{CDP} is defined as a three-tuple, denoted as $\Omega \left \langle t, \mathbf{S}\left ( t \right ), \mho \left ( \cdot  \right ) \right \rangle$, where

\begin{itemize}

\item[*] $t = 0,1,2,\cdots$ is the period index.



\item[*] $\mathbf{S}\left ( t \right ) \in \left \{ 0,1 \right \}^{n\times K}$ is the $n\times K$ community membership matrix during period $t$, where $\mathbf{S}\left ( 0 \right ) = \mathbf{s}\left ( T \right )$



\item[*] $\mho \left ( \cdot  \right )$ is the community dynamics function, which returns the community membership matrix at the next period, i.e., $\mathbf{S}\left ( t+1 \right )$.

\end{itemize}

\end{definition}

During every period $t$, each agent $i$ is associated with a utility function, denoted by
\begin{equation}
	\begin{aligned}
		\label{utility_define}
			u_{i}\left ( t \right ) 
   &= \lambda_{i}\sum_{j \ne i}\mathbf{S}_{i}\left ( t \right )\mathbf{S}_{j}^{\mathsf{T}}\left ( t \right ) \cdot \left[\psi-\left|\mathbf{x}_{i}\left ( T \right ) - \mathbf{x}_{j}\left ( T \right ) \right|\right]\cdot \mathbf{W}_{ij} \\
   &+\sum_{j \ne i}\left ( 1- \mathbf{S}_{i}\left ( t \right )\mathbf{S}_{j}^{\mathsf{T}}\left ( t \right )\right )\cdot ReLu\left[\psi-\left|\mathbf{x}_{i}\left ( T \right ) - \mathbf{x}_{j}\left ( T \right ) \right|\right]\cdot \mathbf{W}_{ij}
	\end{aligned}
\end{equation}

The intuitions behind this definition are: 
(i) Both the opinions of the neighbors and the connections between the agent and the neighbors should be considered; 
(ii) The community labels of neighbors with closer opinions within the confidence level are given greater weight. For simplicity, we assume that all the agents have the same confidence level $\psi$; 
(iii) Agents have a higher level of reliability on the opinions of intra-community neighbors; 
(iv) Excessive differences in opinions among agents within the same community can be detrimental to the stability of the community; 
(v) Agents will only communicate with inter-community neighbors whose opinions are within the confidence level. (Please find the detailed effectiveness analysis in Section \ref{Comparative_discussion})

Let $\mathbf{D}\left ( T \right ) \in \left [ 0, 1 \right ] ^{n\times n}$ denotes the pairwise difference matrix, such that $\mathbf{D}_{ij}\left ( T \right ) = \psi - \left|\mathbf{x}_{i}\left ( T \right )  -\mathbf{x}_{j}\left ( T \right ) \right|$, we can define the following utility vector $\mathbf{U} \left ( t \right )$ to represent the utility of all agents during period $t$

\begin{equation}
	\begin{aligned}
		\label{ao_1}
			\mathbf{U}\left ( t \right ) &= \Lambda \mathbf{S}\left ( t \right )\mathbf{S}^{\mathsf{T}}\left ( t \right ) \odot \mathbf{D}\left ( T \right ) \odot \mathbf{W} + \left ( \mathbf{1}^{n} -\mathbf{S}\left ( t \right )\mathbf{S}^{\mathsf{T}}\left ( t \right ) \right)\odot ReLu\left( \mathbf{D}\left ( T \right ) \right) \odot \mathbf{W}  \\
              &= \mathbf{B}\left ( T \right ) \odot \mathbf{W} + \mathbf{S}\left ( t \right )\mathbf{S}^{\mathsf{T}}\left ( t \right ) \odot \Lambda \mathbf{D}\left ( T \right ) \odot \mathbf{W} - \mathbf{S}\left ( t \right ) \mathbf{S}^{\mathsf{T}}\left ( t \right ) \odot \mathbf{B}\left ( T \right ) \odot \mathbf{W} \\
              &= \mathbf{B}\left ( T \right ) \odot \mathbf{W} +  \mathbf{S}\left ( t \right ) \mathbf{S}^{\mathsf{T}}\left ( t \right ) \odot \left ( \Lambda \mathbf{D}\left ( T \right ) - \mathbf{B}\left ( T \right ) \right ) \odot \mathbf{W},  \\        
	\end{aligned}
\end{equation}

\noindent where $\mathbf{D}\left ( T \right ) = \Psi - \mathbf{E}\left ( T \right )$, $\Psi = \psi \mathbf{1}^{n}$, $\mathbf{E}\left ( T \right ) = \left | \mathbf{x}\left ( T \right ) 1_{n}^{\mathsf{T}} - 1_{n} \mathbf{x}^{\mathsf{T}}\left ( T \right ) \right |$ and $\mathbf{B}\left ( T \right ) = ReLu\left(\mathbf{D}\left ( T \right )  \right )$. Social welfare is defined as the sum of the utilities of all members and can therefore be expressed as

\begin{equation}
	\begin{aligned}
		\label{Social_welfare}
   sw\left ( t \right ) 
   = \sum_{i}u_{i}\left ( t \right ) 
   &= \sum_{ij} \mathbf{U}_{ij}\left ( t \right ) \\
   &= \sum_{ij}\left \{ \left ( \mathbf{S}\left ( t \right ) \mathbf{S}^{\mathsf{T}}\left ( t \right ) \right ) \odot \left ( \Lambda \mathbf{D}\left ( T \right ) - \mathbf{B}\left ( T \right ) \right ) \odot \mathbf{W} \right \}_{ij} + \sum_{ij}\left ( \mathbf{B}\left ( T \right ) \odot \mathbf{W} \right ) _{ij} \\
   &= tr\left \{ \left ( \mathbf{S}\left ( t \right ) \mathbf{S}^{\mathsf{T}}\left ( t \right ) \right )\left [ \left ( \Lambda \mathbf{D}\left ( T \right ) - \mathbf{B}\left ( T \right ) \right ) \odot \mathbf{W} \right ] \right \} + tr\left (  \mathbf{W} \mathbf{B}\left ( T \right ) \right ) \\
   &= tr\left \{\mathbf{S}^{\mathsf{T}}\left ( t \right )\left [ \left ( \Lambda \mathbf{D}\left ( T \right ) - \mathbf{B}\left ( T \right ) \right ) \odot \mathbf{W} \right ] \mathbf{S}\left ( t \right ) \right \} + tr\left (  \mathbf{W} \mathbf{B}\left ( T \right ) \right ),
   \end{aligned}
\end{equation}

\noindent where the first term measures the change of empathetic social welfare with respect to $\mathbf{S}\left ( t \right )$, while the second term, i.e., $tr\left (  \mathbf{W} \mathbf{B}\left ( T \right ) \right )$ only related to neighbors with sufficiently close opinions, and is independent of the community membership matrix $\mathbf{S}\left ( t \right )$. Since \textbf{\emph{CDP}} is primarily concerned with maximizing the social welfare, and $\left ( \Lambda \mathbf{D}\left ( T \right ) - \mathbf{B}\left ( T \right ) \right ) \odot \mathbf{W} $ remains unchanged in the process, the problem that \textbf{\emph{CDP}} addresses can be formally formalized as

\begin{problem}\label{problem:CDSWG} [\textbf{Community Detection for Social Welfare Maximization (CDSWM)}] Given influence matrix $\mathbf{W}$, diagonal matrix $\Lambda$ and  opinion vector $\mathbf{x} \left ( T \right )$, \textbf{CDSWM} aims to find the community membership matrix $\mathbf{S}^{\ast }\in \left \{ 0,1 \right \}^{n\times K}$ that maximizes the social welfare, such that

\begin{equation}
	\begin{aligned}
		\label{problem}
   tr\left \{\left ( \mathbf{S}^{\ast } \right ) ^{\mathsf{T}} \left [ \left ( \Lambda \mathbf{D}\left ( T \right ) - \mathbf{B}\left ( T \right ) \right ) \odot \mathbf{W} \right ] \mathbf{S}^{\ast } \right \} 
   & \ge tr\left \{\mathbf{S}^{\mathsf{T}}\left ( t \right )\left [ \left ( \Lambda \mathbf{D}\left ( T \right ) - \mathbf{B}\left ( T \right ) \right ) \odot \mathbf{W} \right ] \mathbf{S}\left ( t \right ) \right \}, \\
   & t = 0,1,2,\cdots \\
   s.t., \; \; & \forall \mathbf{S}_{i}^{\ast} 1_{K} = 1, \; \; i\in \mathcal{N} 
   \end{aligned}
\end{equation}

\end{problem}

\begin{lemma}\label{lemma:1} The solution of \textbf{CDSWM} is equal to finding the community structure that maximizes $tr\left ( \mathbf{S}^{\mathsf{T}}\left ( t \right ) \mathbf{Y} \left ( T \right ) \mathbf{S}\left ( t \right ) \right )$, where $\mathbf{Y} \left ( T \right ) = \left ( \Lambda \mathbf{D}\left ( T \right ) - \mathbf{B}\left ( T \right ) \right ) \odot \mathbf{W} + \left ( \mathbf{D}^{\mathsf{T}}\left ( T \right ) \Lambda - \mathbf{B}^{\mathsf{T}} \left ( T \right ) \right ) \odot \mathbf{W}^{\mathsf{T}} $ with each entry:

\begin{equation}
	\begin{aligned}
		\label{Y_define}
    \mathbf{Y}_{ij}\left ( T \right ) = \left ( \lambda_{i} \mathbf{D}_{ij}\left ( T \right ) - \mathbf{\mathbf{B}}_{ij}\left ( T \right ) \right ) \mathbf{W}_{ij}+ \left ( \lambda_{j}\mathbf{D}_{ji}\left ( T \right ) - \mathbf{\mathbf{B}}_{ji}\left ( T \right ) \right ) \mathbf{W}_{ji} = \mathbf{Y}_{ji}\left ( T \right )
   \end{aligned}
\end{equation}

\end{lemma}

\noindent\textsc{Proof:} Obviously, $\mathbf{Y} \left ( T \right )$ is the $n\times n$ symmetric matrix  with diagonal elements $\mathbf{Y}_{ii}\left ( T \right ) = 0$. As

\begin{equation}
	\begin{aligned}
		\label{ao_2}
    tr\left \{\mathbf{S}^{\mathsf{T}}\left ( t \right )\left [ \left ( \Lambda \mathbf{D}\left ( T \right ) - \mathbf{B}\left ( T \right ) \right ) \odot \mathbf{W} \right ] \mathbf{S}\left ( t \right ) \right \} 
    = tr\left \{ \mathbf{S}^{\mathsf{T}} \left [ \left ( \mathbf{D}^{\mathsf{T}}\left ( T \right ) \Lambda - \mathbf{B}^{\mathsf{T}} \left ( T \right ) \right ) \odot \mathbf{W}^{\mathsf{T}} \right ] \mathbf{S}\left ( t \right ) \right \},
   \end{aligned}
\end{equation}

\noindent we have 

\begin{equation}
	\begin{aligned}
		\label{ao_2}
    & tr\left ( \mathbf{S}^{\mathsf{T}}\left ( t \right ) \mathbf{Y} \left ( T \right ) \mathbf{S}\left ( t \right ) \right ) \\
    &= tr\left \{\mathbf{S}^{\mathsf{T}}\left ( t \right )\left [ \left ( \Lambda \mathbf{D}\left ( T \right ) - \mathbf{B}\left ( T \right ) \right ) \odot \mathbf{W} + \left ( \mathbf{D}^{\mathsf{T}} \left ( T \right ) \Lambda - \mathbf{B}^{\mathsf{T}} \left ( T \right ) \right ) \odot \mathbf{W}^{\mathsf{T}} \right ] \mathbf{S}\left ( t \right ) \right \} \\
    &= 2tr\left \{\mathbf{S}^{\mathsf{T}}\left ( t \right )\left [ \left ( \Lambda \mathbf{D}\left ( T \right ) - \mathbf{B}\left ( T \right ) \right ) \odot \mathbf{W} \right ] \mathbf{S}\left ( t \right ) \right \},
   \end{aligned}
\end{equation}

\noindent which suggests that the community membership matrix $\mathbf{S}^{\ast }$ that maximizes $tr\left ( \mathbf{S}^{\mathsf{T}}\left ( t \right ) \mathbf{Y} \left ( t \right ) \mathbf{S}\left ( t \right ) \right )$ can also maximize $tr\left \{\mathbf{S}^{\mathsf{T}}\left ( t \right )\left [ \left ( \Lambda \mathbf{D}\left ( T \right ) - \mathbf{B}\left ( T \right ) \right ) \odot \mathbf{W} \right ] \mathbf{S}\left ( t \right ) \right \}$. In this context, we can define the following non-cooperative game to find an optimal solution to \textbf{\emph{CDSWM}}:

\begin{definition}\label{def:CFG} [\textbf{dynamics Game for Social Welfare Maximization (DGSWM).}] The \textbf{DGSWM} is defined as a non-cooperative game $\Pi\left ( t \right ) =  \left \langle \mathcal{S}\left ( t \right ), u^{\star}\left ( \cdot \right ), \Gamma \left ( \mathbf{S} \left ( t \right )  \right ) \right \rangle $ with agents are assumed to operate on the principle of utility maximization, where

\begin{itemize}


\item[*] $\mathcal{S}\left ( t \right )= \left \{ \mathcal{S}_{1}\left ( t \right ), \mathcal{S}_{2}\left ( t \right ),\dots , \mathcal{S}_{n}\left ( t \right ) \right \}$ is the combination of agents’ strategies, where 

\begin{equation}
	\begin{aligned}
		\label{modify_utility}
    \mathcal{S}_{i}\left ( t \right ) = \left \{ e_{\left ( k \right )}^{\mathsf{T}}: \mathbf{W}_{ij} + \mathbf{W}_{ji}> 0, \mathbf{S}_{j} \left ( t \right ) = e_{\left ( k \right )}^{\mathsf{T}}\right \} \cup \left \{ \mathbf{S}_{i} \left ( t \right ) \right \}
   \end{aligned}
\end{equation}

is the finite set of community labels that agent $i$ can choose from.

\item[*] $u^{\star}\left ( \mathbf{S} \left ( t \right ) \right ): \mathbf{S}_{1} \times \mathbf{S}_{2} \times \cdots \times \mathbf{S}_{n} \to  \mathbb{R} $ is used to calculate the modified utility associated with agents under a given community structure, where the utility of any agent $i$ is

\begin{equation}
	\begin{aligned}
		\label{modify_utility}
    u_{i} ^{\ast }  \left ( \mathtt{S}\left ( t \right ) \right ) = \sum_{j\ne i}\mathtt{S}_{i} \left ( t \right ) \mathtt{S}_{j}^{\mathsf{T}}\left ( t \right ) \mathbf{Y}_{ij}\left ( t \right )
   \end{aligned}
\end{equation}

To speed up the update process, we only consider that $\mathtt{S}_{i} \left ( t \right ) \in \mathcal{S}_{i}\left ( t \right )$ because $ \forall e_{\left ( \pi \right )}^{\mathsf{T}} \notin \mathcal{S}_{i}\left ( t \right ) \Rightarrow u_{i}^{\ast } \left ( \left [ e_{\left ( \pi \right )}^{\mathsf{T}},\mathtt{S}_{-i} \right ]  \right ) =  {\textstyle \sum_{j\in \mathcal{I}_{i} }}\left ( e_{\left ( \pi \right )}^{\mathsf{T}}\mathtt{S}_{j}^{\mathsf{T}}\mathbf{Y}_{ij} \left ( t \right ) \right ) = 0$, where $\mathcal{I}_{i} = \left \{ j \mid \mathbf{W}_{ij} + \mathbf{W}_{ji}> 0  \right \}$.

\item[*] $\Gamma \left ( \mathbf{S} \left ( t \right )  \right ): \mathbf{S} \to \mathbf{S}$ is the transition function that returns the next stage of each agent's community label based on the utility calculated by the current community structure. For any agent $i$,
\begin{equation}
	\begin{aligned}
		\label{modify_utility}
    \mathbf{S}_{i} \left ( t + 1 \right )  =  \Gamma_{i} \left ( \mathbf{S} \left ( t \right )  \right ) = \underset{ \mathtt{S}_{i}^{\star  } \in \mathcal{S}_{i}\left ( t \right )}{argmax} \enspace u_{i} ^{\ast }  \left ( \mathtt{S}_{i}^{\star  }, \mathbf{S}_{- i}\left ( t \right ) \right )
   \end{aligned}
\end{equation}

\end{itemize}

\end{definition}

\begin{theorem}\label{theorem:1} The \textbf{DGSWM} introduced admits a pure Nash equilibrium \cite{pure-nash} because there exists a potential function

\begin{equation}
	\begin{aligned}
		\label{potential_function}
    \phi ^{\ast }  \left ( \mathbf{S} \right ) = \frac{1}{2} \sum_{ij}\mathbf{S}_{i}\mathbf{S}_{j}^{\mathsf{T}}\mathbf{Y}_{ij} = \frac{1}{2} tr(\mathbf{S}^{\mathsf{T}} \mathbf{Y} \mathbf{S})=tr(\mathbf{S}^{\mathsf{T}} \mathbf{Z} \mathbf{S}),
   \end{aligned}
\end{equation}

\noindent such that $\phi^{\ast}\left ( \left [ \mathbf{S}_{i}^{\prime },\mathbf{S}_{-i} \right ] \right ) -\phi^{\ast}\left ( \left [ \mathbf{S}_{i},\mathbf{S}_{-i} \right ] \right ) = u_{i}^{\ast }\left ( \left [ \mathbf{S}_{i}^{\prime },\mathbf{S}_{-i} \right ]  \right ) -u_{i}^{\ast }\left ( \left [ \mathbf{S}_{i},\mathbf{S}_{-i} \right ]  \right )$.

\end{theorem}

(See \ref{de:Feature_DGSWG} for proof)

Finally, we will update the agents' strategies ($\mho \left ( \cdot  \right )$ in \textbf{\emph{CDP}}) with the \emph{best-response dynamics} \cite{GLEAM} (Line [11] to Line [20], Algorithm \ref{alg}). For $n$ agents, we randomly select a agent $i$ based on sampling without replacement at a time, and keep the strategies of the other agents unchanged, then use the transition function $\Gamma_{i} \left ( \mathbf{S} \left ( t \right )  \right )$ to update the community label of $i$. Each iteration process requires updating all agents, and if the community labels of all agents does not change during an iteration, a pure Nash equilibrium has been found.

\begin{theorem}\label{theorem:2} Given any community member matrix $\mathbf{s}\left ( T \right )$ at the initial moment, \textbf{\emph{CDP}} can ultimately find a stable community membership matrix $\mathbf{s}\left ( T + 1\right )$, which is a pure Nash equilibrium w.r.t. \textbf{\emph{DGSWM}}, and is also the solution of \textbf{\emph{CDSWM}}.
\end{theorem}

(See \ref{de:Effectiveness_CDP} for proof)

\begin{theorem}\label{theorem:4} Given that there are $M$ directed connections among agents, from arbitrary initial opinion vector $\mathbf{x} \left ( T \right )$ and community membership matrix $\mathbf{s} \left ( T \right )$, the time complexity of CDP is upper-bounded by $\mathcal{O} \left ( t^{Exp} M \right )$, where $t^{Exp}$ is the expected number of iterations required by the \emph{best-response dynamics} to reach the termination condition.
\end{theorem}

(See \ref{de:Complexity_CDP} for proof)

\subsection{Opinion Formation Process}


During the opinion update step, we take into account the disparities in opinions among the agents as well as the impact of the community structure in the present time. Agents exhibit a greater degree of dependability when relying on the viewpoints of their fellow community members, while disregarding the opinions of neighbors beyond a certain level of confidence. The opinion update rule for agents can be described as follows.

\begin{definition}\label{def:CAOFP} [\textbf{Community-Aware Opinion Formation Process (CAOFP)}] Once the community membership matrix has been updated, each agent will update her opinion using the received information from neighbors, and the dynamics of opinions are given by

\begin{equation}
	\begin{aligned}
		\label{opin_dynamic}
    \mathbf{x}_{i} \left ( T+1 \right ) &= \delta_{i} \left ( T \right ) \mathbf{x}_{i}\left ( T \right ) + \sum_{j\ne i}\Phi_{ij}\left ( T \right ) \cdot \mathbf{x}_{j}\left ( T \right ),
   \end{aligned}
\end{equation}

\end{definition}

\noindent where the opinion weight $\Phi_{ij}\left ( T \right )$ that agent $i$ assigns to $j$ at time $T$ can be defined as

\begin{equation}
	\begin{aligned}
		\label{ao_2}
    \Phi_{ij}\left ( T \right )
    &= \frac{\lambda_{i} \mathbf{s}_{i}\left ( T \right ) \mathbf{s}_{j}^{\mathsf{T}}\left ( T \right ) \omega_{ij} +\left [ 1-\mathbf{s}_{i}\left ( T \right ) \mathbf{s}_{j}^{\mathsf{T}}\left ( T \right ) \right ] \omega_{ij}  }{\lambda _{i}\sum_{j\ne i}\mathbf{s}_{i}\left ( T \right ) \mathbf{s}_{j}^{\mathsf{T}}\left ( T \right ) \omega_{ij} + \sum_{j\ne i}\left [ 1-\mathbf{s}_{i}\left ( T \right ) \mathbf{s}_{j}^{\mathsf{T}}\left ( T \right ) \right ] \omega_{ij} } \\
    &= \frac{ \left [ 1 + \left( \lambda_{i} - 1 \right ) \mathbf{s}_{i}\left ( T \right ) \mathbf{s}_{j}^{\mathsf{T}}\left ( T \right ) \right ] \omega_{ij}  }{\sum_{j\ne i} \left [ 1 + \left( \lambda_{i} - 1 \right ) \mathbf{s}_{i}\left ( T \right ) \mathbf{s}_{j}^{\mathsf{T}}\left ( T \right ) \right ] \omega_{ij} + 1 } \\
    s.t., \omega_{ij} &= \overbrace{ \mathbf{W}_{ij}\cdot Relu\left [ \beta \gamma ^{T} - \left|\mathbf{x}_{i}\left ( T \right ) - \mathbf{x}_{j}\left ( T \right ) \right| \right] }^{j\in \mathcal{N}_{i}^{T} }
   \end{aligned}
\end{equation}

    
    

\noindent $\mathcal{N}_{i}^{T}= \left \{ j\mid j\in N_{i} \wedge \left|\mathbf{x}_{i}\left ( T \right ) - \mathbf{x}_{j}\left ( T \right ) \right | < \beta \gamma ^{T} \right \}$ denotes the set of the trusted neighbors of user $i$ at time $T$. $\beta\in \left ( 0,1 \right ) ,\gamma\in \left ( 0,1 \right ) $ are the tuning
parameters for the size of $\mathcal{N}_{i}^{T}$. Obviously, as the number of iterations increases, the agents' opinions will gradually be consolidated and their confidence level will continue to decrease. The opinion weight matrix during time $T$, denoted by $\Phi \left ( T \right )$, can be written as

\begin{equation}
	\begin{aligned}
		\label{ao_2}
    \Phi \left ( T \right ) = \Theta\left ( T \right ) \oslash \left [  \Theta\left ( T \right ) \mathbf{1}^{n} + \mathbf{1}^{n} \right ], \\
   \end{aligned}
\end{equation}

\begin{equation}
	\begin{aligned}
		\label{ao_2}
    \Theta\left ( T \right ) = \left [ \mathbf{1}^{n} +\left ( \Lambda - \mathbf{1}_{n} \right ) \mathbf{s}\left ( T \right ) \mathbf{s}^{\mathsf{T}}\left ( T \right ) \right ]  \odot \mathbf{W} \odot Relu \left ( \beta \gamma ^{T} \mathbf{1}^{n} - \mathbf{E} \left ( T \right ) \right ), \\
   \end{aligned}
\end{equation}

\noindent where $\oslash$ means element-wise divide, i.e., $\Phi_{ij}\left ( T \right )  = \Theta_{ij}\left ( T \right ) {\div} \left ( \sum_{j\ne i} \Theta_{ij}\left ( T \right ) + 1 \right )$. During every time $T$, each agent $i$ is also associated with a intrinsic confidence level, denoted by $\delta_{i} \left (  T \right )$, which is defined as

\begin{equation}
	\begin{aligned}
		\label{ao_2}
    \delta_{i}\left ( T \right ) = \frac{1 }{\sum_{j\ne i} \left [ 1 + \left( \lambda_{i} - 1 \right ) \mathbf{s}_{i}\left ( T \right ) \mathbf{s}_{j}^{\mathsf{T}}\left ( T \right ) \right ]\cdot \mathbf{W}_{ij}\cdot Relu\left [ \beta \gamma ^{T} -\left|\mathbf{x}_{i}\left ( T \right ) - \mathbf{x}_{j}\left ( T \right ) \right| \right] + 1 } \\
   \end{aligned}
\end{equation}

Therefore, the opinion update rule Eq.~(\ref{opin_dynamic}) can be written in matrix form as
\begin{equation}
	\begin{aligned}
		\label{ao_2}
    \mathbf{x}\left ( T+1 \right ) = 1_{n} \oslash \left [ \Theta\left ( T \right ) 1_{n} + 1_{n} \right ] \odot \mathbf{x}\left ( T \right ) +  \Phi \left ( T \right ) \mathbf{x}\left ( T \right )   \\
   \end{aligned}
\end{equation}

\begin{lemma}\label{lemma:2} $\mathbf{x}\left ( T+1 \right )$ must be in the interval of $\left [ 0, 1 \right ]$, as $\forall i\in \mathcal{N}$, $\delta_{i}\left( T \right) + \sum_{j \ne i}\Phi_{ij}\left( T \right) = 1$.
\end{lemma}

\begin{remark}\label{remark:1} \textbf{CAOFP} introduces a community mechanism into the opinion update rule, which provides a feasible scheme for modeling the co-evolution of communities and opinions. In addition, communities are formed naturally and systematically in a bottom-up manner in \textbf{CDP}, so prior information such as the number of communities is not required. Therefore, GCAOFP can automatically identify the underlying community structures in social networks and construct the corresponding dynamics evolution processes of opinions.
\end{remark}

\begin{lemma}\label{lemma:3} The proposed model degenerates into the \emph{Hegselmann-Krause} model \cite{HK} when the following conditions are met:

\begin{enumerate} 

\item {During each time $T$, $\forall i\in \mathcal{N}, \mathbf{s}_{i}\left ( T \right ) = 1_{K}^{\mathsf{T}}$, i.e., $\forall i,j \in \mathcal{N}, \mathbf{s}_{i}\left ( T \right ) \mathbf{s}_{j}^{\mathsf{T}}\left ( T \right ) = 1$, implying that ignores the community structure hidden in the social network.}

\item {$\Lambda  = \mathbf{1}_{n}$, means each agent trust the the agents in other communities as interactive agents in the same community.}

\item {$\beta \gamma ^{T} -\left | \mathbf{x}_{i}\left ( T \right )- \mathbf{x}_{j}\left ( T \right )\right | = 1$ if $\left | \mathbf{x}_{i}\left ( T \right )- \mathbf{x}_{j}\left ( T \right )\right | < \beta \gamma ^{T}$}; otherwise equal to 0.

\item {$\gamma  = 1$}, i.e., the confidence level of the agent remains unchanged during the update process.

\end{enumerate}
\end{lemma}

(See \ref{de:degenerates_hk} for proof)

\begin{theorem}\label{theorem:5} If the opinion dynamics follows the function in Eq.~(\ref{opin_dynamic}), for every user $i$ and every time $T \ge 0$, we have
\begin{equation}
	\begin{aligned}
		\label{ao_2}
    \left | \mathbf{x}_{i}\left ( T+1 \right ) - \mathbf{x}_{i}\left ( T \right ) \right | < \beta \gamma ^{T} \\
   \end{aligned}
\end{equation}
\end{theorem}

(See \ref{de:Convergence_of_OFP} for proof)

Therefore, we claim that every agents $i$’s opinion vector will converge to a relatively stable state in finite periods $T^{\ast }$, such that $ \forall i \in N , \left | \mathbf{x}_{i}\left ( T^{\ast } \right ) - \mathbf{x}_{i}\left ( T^{\ast }-1 \right ) \right | < \epsilon $, where $\epsilon$ is some very small positive threshold. This is also the termination condition of Algorithm \ref{alg} (Line [27]) .

\begin{theorem}\label{theorem:6} From any arbitrary initial distribution of opinions $\mathbf{x} \left ( 0 \right )$ and any arbitrary initial community membership matrix $ \mathbf{s} \left ( 0 \right )$, the \emph{CDP} introduced in Definition \ref{def:CDP} will converge to the equilibrium state where the community membership matrix keeps unchanged.
\end{theorem}

(See \ref{de:Convergence_of_Community} for proof)

\begin{theorem}\label{theorem:7} Given that there are $M$ directed connections among agents, from arbitrary initial opinion vector $\mathbf{x} \left ( 0 \right )$ and community membership matrix $\mathbf{s} \left ( 0 \right )$, the time complexity of GCAOFP introduced in Algorithm \ref{alg} is upper-bounded by $\mathcal{O} \left ( t^{Exp} M \right )$.
\end{theorem}

(See \ref{de:Complexity_GCAOFP} for proof)

\subsection{Comparative Discussion}\label{Comparative_discussion}


\begin{wrapfigure}{l}{7cm}
    \centering
    \includegraphics[width=1.1in]{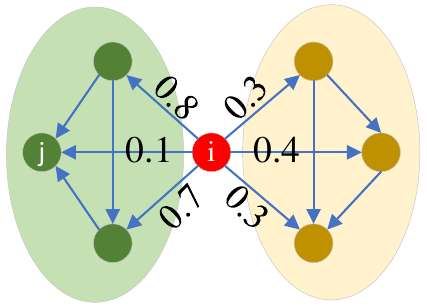}
    \hspace{0.1in}
	\includegraphics[width=1.1in]{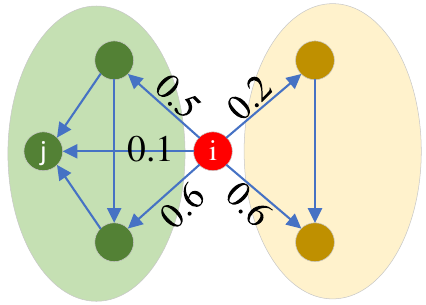}
    \caption{(Left): The illustration of the difference between \emph{GCAOFP} and \emph{LPA-HK} on candidate strategy selection during the community dynamics process; (Right): Demonstration of the rationality of introducing the influence of inter-community neighbors into Eq.~(\ref{utility_define}). We assume that the connections among agents are unweighted. The pairwise $\mathbf{E}_{ij}$ values associated with agent $i$ are marked in numbers, and the colors of nodes indicate their community labels.}
    \label{fig: advertisement}
\end{wrapfigure}

The \emph{GK-Means} algorithm establishes the utility of an agent by quantifying the disparity between their own opinion and the collective opinion of the community. This evaluation occurs during the search for a locally Pareto-optimal community structure. Subsequently, the agent proceeds to update their community label by selecting a feasible strategy that maximizes their utility from within their own set of options. The \emph{GK-Means} algorithm is designed to assess the opinion of an individual agent by comparing it to the average opinions of communities. This allows for the determination of the community preferences of the agent. On the other hand, the \emph{GCAOFP} approach adopts a micro-perspective and argues that the community preference of each agent should be decided based on the opinions of her neighbors. Our technique demonstrates superior efficiency and effectiveness due to the inherent limitations of individuals in perceiving just local information while lacking the ability to comprehend the broader global information of society.


The \emph{LPA-HK} algorithm exclusively takes into account the connections and opinions among agents during the community label update stage. It assigns higher importance to the labels of neighbors who have similar ideas within the confidence level. The relevant utility function can be expressed as follows.

\begin{equation}
	\begin{aligned}
		\label{LPA-HK-UTILITY}
    u_{i}^{\prime} \left ( T \right )  = \sum_{j\ne i} \mathbf{s}_{i}\left ( T \right ) \left ( \mathbf{s}^{\ast  }_{j} \left ( T \right ) \right ) ^{\mathsf{T}}  \varepsilon \left ( \psi -\left | \mathbf{x}_{i}\left ( T \right ) - \mathbf{x}_{j}^{\ast } \left ( T \right ) \right |  \right ) \left ( 1 -\left | \mathbf{x}_{i}\left ( T \right ) - \mathbf{x}_{j}^{\ast } \left ( T \right ) \right |  \right ) \mathbf{W}_{ij},
   \end{aligned}
\end{equation}

Can be implemented in a simpler way: 
\begin{equation}
	\begin{aligned}
		\label{LPA-HK-UTILITY}
    u_{i}^{\prime} \left ( T \right )  = \sum_{j\ne i} \mathbf{s}_{i} \left ( T \right ) \left ( \mathbf{s}^{\ast  }_{j} \left ( T \right ) \right ) ^{\mathsf{T}}  ReLu \left ( \psi -\left | \mathbf{x}_{i}\left ( T \right ) - \mathbf{x}_{j}^{\ast } \left ( T \right ) \right | \right ) \mathbf{W}_{ij},
   \end{aligned}
\end{equation}

\noindent where $\mathbf{s}_{j}^{\ast } \left ( T \right ) = \mathbf{s}_{j} \left ( T \right )$ and $\mathbf{x}_{j}^{\ast } \left ( T \right ) = \mathbf{x}_{j} \left ( T \right )$ if the opinion and label of agent $j$ have already been updated, otherwise $\mathbf{s}_{j}^{\ast } \left ( T \right ) = \mathbf{s}_{j} \left ( T -1 \right )$ and $\mathbf{x}_{j}^{\ast } \left ( T \right ) = \mathbf{x}_{j} \left ( T -1 \right )$.


Obviously, Eq.~(\ref{LPA-HK-UTILITY}) only considers neighbors within the confidence level, which poses the challenge that the choice of candidate community labels by individual agent can be easily dominated by the neighbor who has the closest opinions to her. Fig.\ref{fig: advertisement} (Left) shows an example of this extreme case. If $\psi = 0.4$, according to Eq.~(\ref{LPA-HK-UTILITY}), agent $i$ will choose to join the left community, even if the opinions of all agents except agent $j$ in the community are outside the confidence level. Eq.~(\ref{utility_define}) perfectly addresses this potential flaw by imposing penalties on intra-community neighbors whose opinions outside the confidence level. With Eq.~(\ref{utility_define}), agent $i$ will join the right community because the utility obtained by agent $i$ joining the left community is $-0.4$, while joining the right community is $0.60$ ($\lambda = 1.5$). Another advantage of Eq.~(\ref{utility_define}) is that it comprehensively considers the influence of inter-community neighbors, which further ensures that agents with significant differences in opinion can be divided into different communities. Taking Fig.\ref{fig: advertisement} (Right) as an example, agent $i$ will ultimately choose to join the right community because although the intra-community utilities ($\lambda_{i}\sum_{j \ne i}\mathbf{S}_{i}\left ( t \right )\mathbf{S}_{j}^{\mathsf{T}}\left ( t \right ) \cdot \left[\psi-\left|\mathbf{x}_{i}\left ( T \right ) - \mathbf{x}_{j}\left ( T \right ) \right|\right]\cdot \mathbf{W}_{ij}$) that agent $i$ can obtain after joining both communities are 0, the inter-community utility ($\sum_{j \ne i}\left ( 1- \mathbf{S}_{i}\left ( t \right )\mathbf{S}_{j}^{\mathsf{T}}\left ( t \right )\right )\cdot ReLu\left[\psi-\left|\mathbf{x}_{i}\left ( T \right ) - \mathbf{x}_{j}\left ( T \right ) \right|\right]\cdot \mathbf{W}_{ij}$) that agent $i$ can obtain after joining the right community is slightly greater ($0.3 > 0.2$). Therefore, compared to \emph{LPA-HK}, the opinions among members within the communities divided by \emph{GACOFP} are closer.

\section{Experiments}

In this section, we will conduct comprehensive experiments using real-world networks to verify the \emph{effectiveness}, \emph{scalability} and \emph{parameter sensitivity} of \emph{GCAOFP}.

\subsection{Experiment Setup.}

\subsubsection{Comparison Methods.} In this work, we will compare \emph{GCAOFP} with some classic standard opinion dynamics models (\emph{DeGroot model} (\emph{DeGroot}) \cite{DGROOT}, \emph{Friedkin-Johnsen model} (\emph{FJ}) \cite{FJ}, \emph{DeGroot-Friedkin model} (\emph{DF}) \cite{DF-MODEL} and \emph{Hegselmann-Krause model} (\emph{HK}) \cite{HK}), which do not consider the impact of community structure in networks. In addition, we have also selected two recently proposed community-related dynamics models \emph{GK-Means} \cite{GK-MEANS} and \emph{LPA-HK} \cite{LPA-HK}, where opinions and community structures can influence each other and co-evolve. Unless noted, for \emph{FJ}, \emph{DF}, and \emph{GK-Means}, we assume that the susceptibility of all agents to social influences are 0.5, thus set the susceptibility matrix $\Lambda = 0.5 \mathbf{1}_{n}$; for \emph{HK} and \emph{LPA-HK}, the range of confidence $\delta$ is set to 0.8.

\subsubsection{Evaluation Metrics.} We define evaluation metrics \emph{OSWG} to measure the ability of \emph{GCAOFP} to improve social welfare with the given initial situation, \emph{ICSW} and \emph{RCSW} to quantify the level of consensus reached among agents at the initial and eventual steady states, respectively. And finally employ \emph{ACL} to reflect the level of consensus among members within the identified communities. (See \ref{de:Evaluation_Metrics_NEW} for detailed definition)


\subsubsection{Network Datasets.} The eight real-world networks shown in Table \ref{tab:dataset} were utilized for the purpose of conducting instance analysis. In order to obtain experimental data from raw network data, a preliminary step involves the removal of any self-loops that may be present. Subsequently, the largest strongly connected components are extracted. The availability of all data sets is public, and we have also furnished processed data.

\subsubsection{Repeatability.} For ease of analysis, we assume that each agent has the same susceptibility to their community labels, i.e., $\lambda_{1}= \lambda_{2}= \dots \lambda_{n}=\lambda$. Furthermore, in all experiments, the tuning parameters $\beta$ and $\gamma$ are set to 0.9 and 1, respectively. For each agent $i$, we randomly select a real number $\chi$ from the interval $\left [ 0, 1 \right ]$ as the initial opinion ($\mathbf{x}_{i}\left ( 0 \right ) = \chi$) and an integer $\kappa$ from the interval $\left [ 0, K-1 \right ]$ as the initial community label ($\mathbf{s}_{i \kappa }\left ( 0 \right ) = 1$). We have released the source code of the proposed algorithm on \href{https://github.com/ZINUX1998/GCAOFP}{\texttt{https://github.com/ZINUX1998/GCAOFP}}.

\begin{table}[t]
\centering \caption{Statistics of (extracted) largest strongly connected components of real-world networks.}
\label{tab:dataset}
\begin{tabular}{ccccccc}
\toprule
Network & Directed & Weighted &\# Nodes & \# Edges & $\langle d \rangle$ & \# GT Communities\\

\midrule

\href{http://www-personal.umich.edu/~mejn/netdata/}{\texttt{Karate}}
& $\times$ & $\times$
& 34 & 78 & 4.59 & 2\\

\href{https://networkrepository.com/soc-dolphins.php}{\texttt{Dolphins}}
&$\times$ & $\times$
&62&159&5.13&2\\

\href{http://www.orgnet.com/divided.html}{\texttt{PolBooks}}
&$\times$  & $\times$
&105&441&8.40&3\\

\href{http://www-personal.umich.edu/~mejn/netdata/}{\texttt{PolBlogs}}
&\Checkmark & $\times$
&793&15,783&19.90&2\\

\midrule

\href{https://snap.stanford.edu/data/soc-sign-bitcoin-alpha.html}{\texttt{BitAlpha}}
&\Checkmark & \Checkmark
&3,235&23,299&7.20&$\times$\\

\href{https://snap.stanford.edu/data/soc-sign-bitcoin-otc.html}{\texttt{BitOTC}}
&\Checkmark  & \Checkmark
&4,709&33,461&7.11&$\times$\\

\href{https://snap.stanford.edu/data/ca-GrQc.html}{\texttt{GRQC}}
&$\times$ & $\times$
&4,158&13,425&6.58&$\times$\\

\href{https://snap.stanford.edu/data/ca-HepTh.html}{\texttt{HEPH}}
&$\times$ & $\times$
&8,638&24,806 &5.74&$\times$\\

\bottomrule

\end{tabular}
\end{table}

\subsection{Experimental Results.}

\subsubsection{Effectiveness Analysis.}\label{sec:analysis}


In order to assess the efficacy of \emph{GCAOFP}, a total of 100 randomly generated initial opinion vectors and community member matrices were applied to each network shown in Table \ref{tab:dataset}.

\begin{wrapfigure}{l}{7cm}
\center
\includegraphics[width=0.5\textwidth]{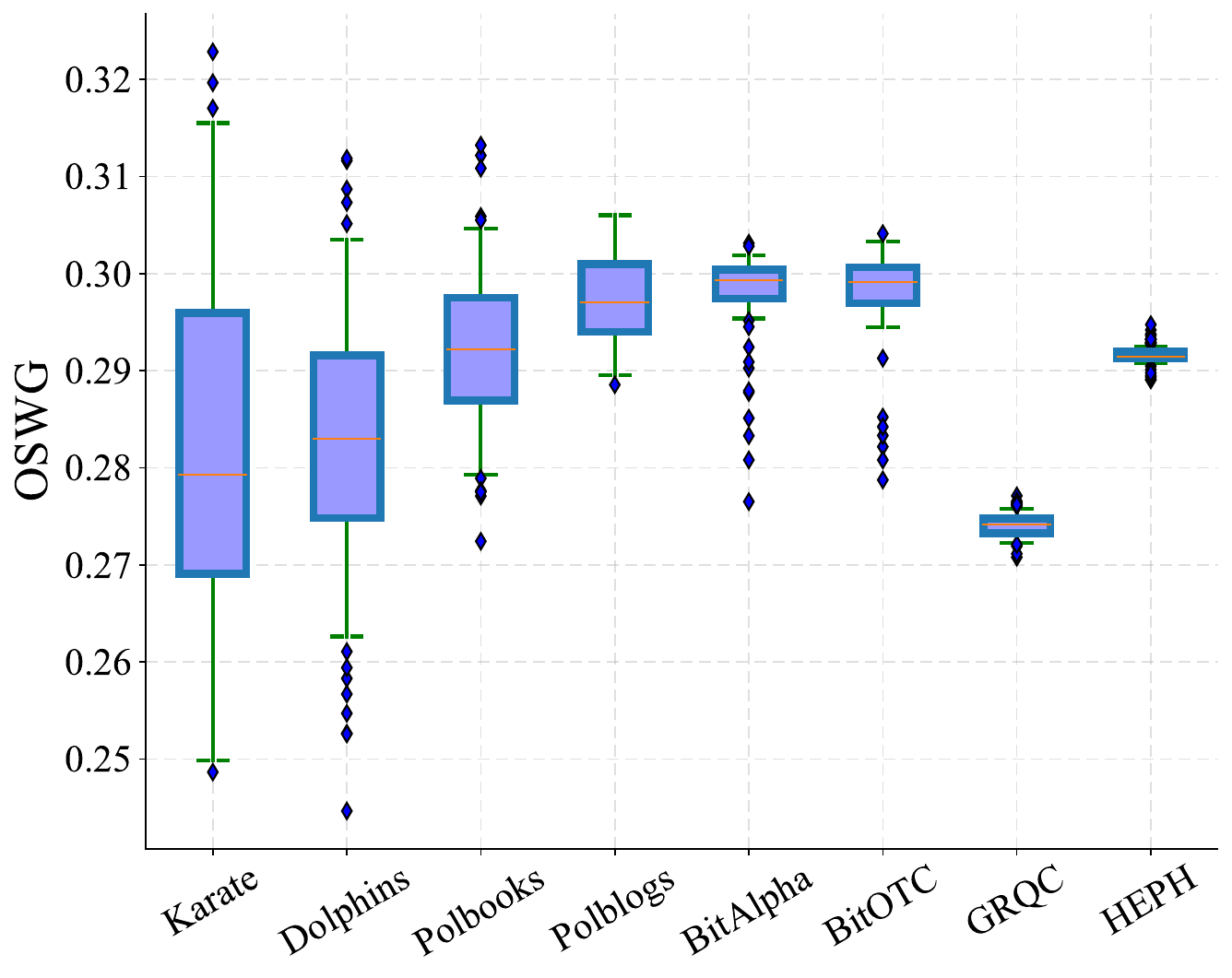}
\caption{The boxplots of \emph{OSWG} (100 implements) of \emph{GCAOFP} on real-world networks. ($\lambda=1.4$ and $\psi=0.4$)}
\label{fig:OSWG-RCSW}
\end{wrapfigure}

Fig.\ref{fig:OSWG-RCSW} presents a summary of the outcomes obtained from the \emph{OSWG}. It is evident from the figure that the proposed \emph{GCAOFP} has the ability to considerably enhance the initial social welfare, regardless of the initial conditions. Another notable observation is that increasing the density of the experimental network has the potential to decrease the variance of \emph{OSWG}. This can be attributed to the fact that a higher average degree $\langle d \rangle$ allows for wider channels for the dissemination of agents' opinions. Consequently, the range of opinions that individual agents can receive becomes more diverse, leading to an enhancement in the resilience of \emph{GCAOFP} to the initial configuration. \ref{de:density_robustness} of the study offers additional confirmation through the utilization of synthetic networks.

\begin{figure}[htbp]
	\centering
	\begin{minipage}{0.99\linewidth}
		\centering
		\includegraphics[width=0.99\linewidth]{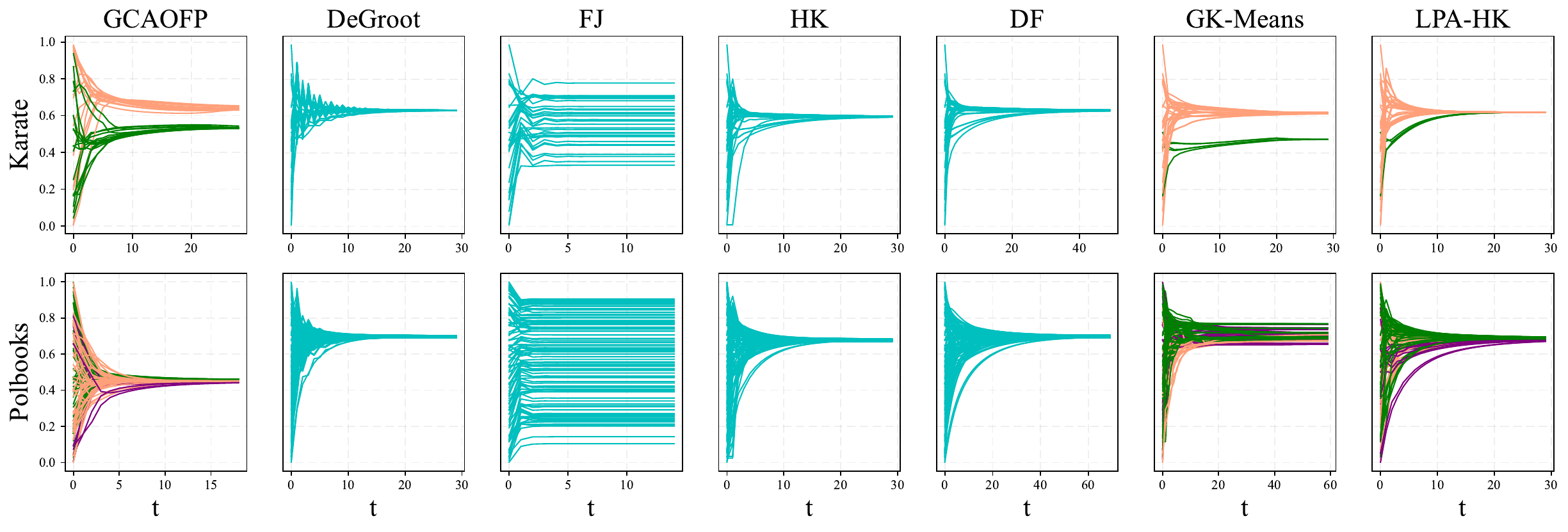}
		\caption{The dynamics of agents' opinions of different dynamics models on \emph{Karate} and \emph{PolBooks}. The color of each opinion curve in each subplot indicates the community label of the corresponding agent obtained using different dynamics models.}
		\label{fig:OPINION_DYNAMIC}
	\end{minipage}
	\qquad
	\begin{minipage}{0.99\linewidth}
		\centering
		\includegraphics[width=0.99\linewidth]{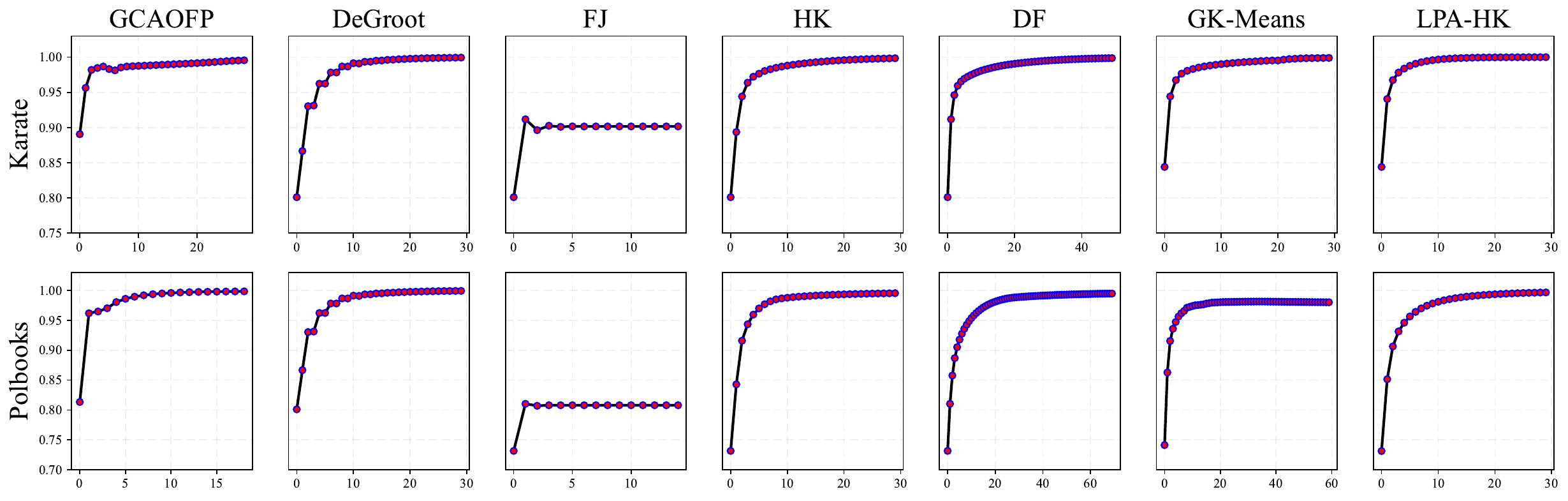}
		\caption{The dynamics of \emph{ACL} of different dynamics models on \emph{Karate} and \emph{PolBooks}.}
		\label{fig:ACL_KARATE_POLBOOKS}  
    \end{minipage}
    \qquad
    \begin{minipage}{0.99\linewidth}
		\centering
		\includegraphics[width=0.99\linewidth]{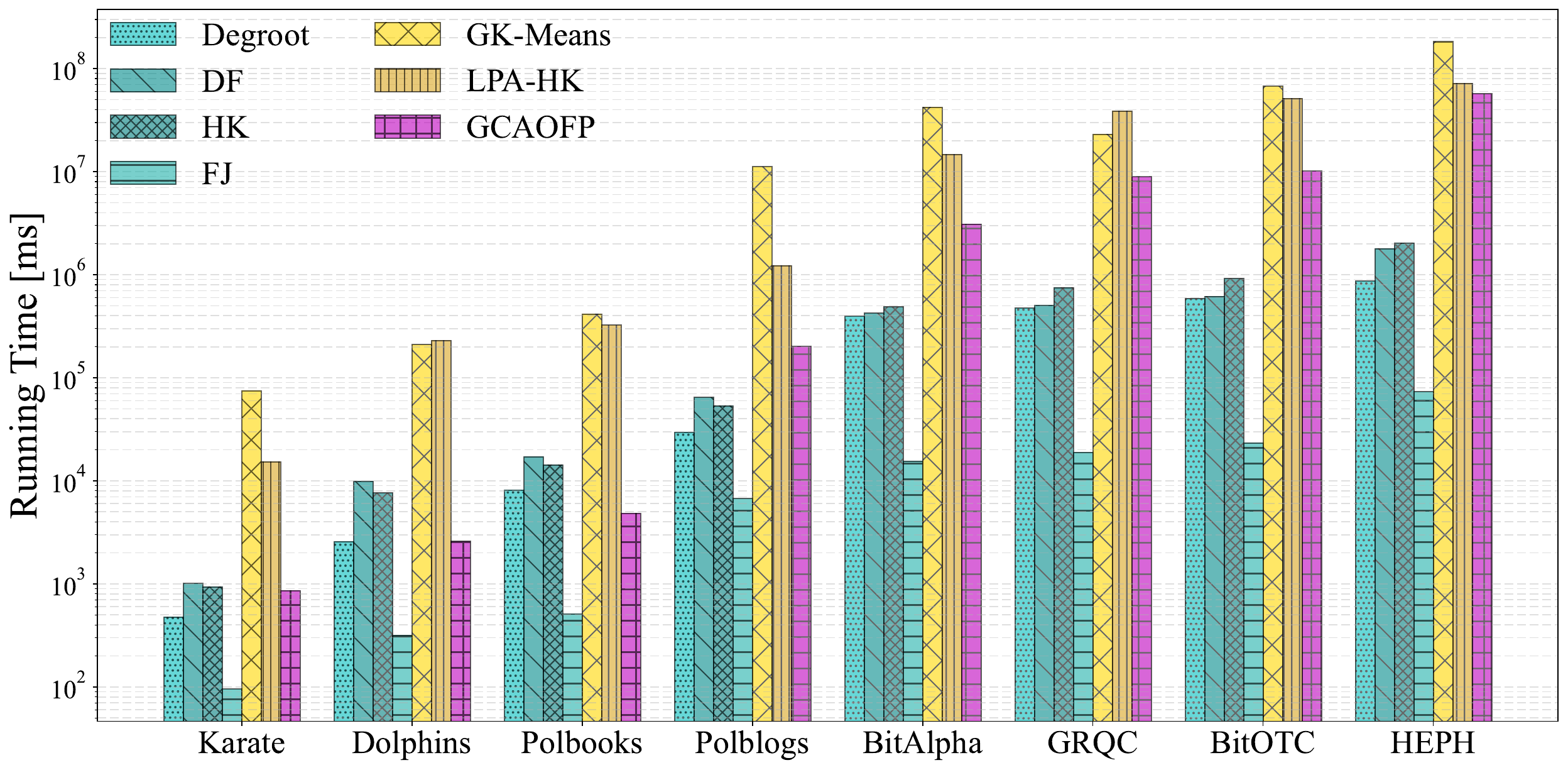}
		\caption{The running time of different dynamics models with increasing number of links (Sorted by the number of links).}
		\label{fig:RUN_TIMES}  
    \end{minipage}
\end{figure}

In this study, we proceed to apply various dynamics models to the \emph{Karate} and \emph{PolBooks} datasets. We then illustrate the progression of each agent's viewpoint and the related evolution of the \textbf{\emph{A}}verage \textbf{\emph{C}}onsensus \textbf{\emph{L}}evel (\emph{ACL}) in Fig.\ref{fig:OPINION_DYNAMIC} and Fig.\ref{fig:ACL_KARATE_POLBOOKS}, respectively, for a single implementation. Based on the evidence presented, it can be inferred that

\begin{enumerate}[1)]

\item The convergence speed of \emph{GCAOFP} is quite satisfactory, and can achieve convergence within similar iterations as other models;

\item With the exception of \emph{FJ}, all standard opinion dynamics models can guarantee that the opinions of agents converge to consensus, i.e., their \emph{ACL} values will evolve to 1;

\item The most important rationality of \emph{GCAOFP} is that the convergence of opinions occurs within the same community, which allows the emergence of multiple opinions in the network, and also ensures every agent's opinion converge to consensus locally. Based on the equilibrium opinions learned by \emph{GCAOFP}, the community partition of the networks can be clearly distinguished. We can observe that there is a clear demarcation among the opinions of agents belong to different communities divided by \emph{GCAOFP}, and the opinions of agents within the same community can always reach consensus. It is also worth pointing out that the quality of the community divided by \emph{GCAOFP} is obviously better than that of other models, especially on \emph{Karate}, where \emph{GCAOFP} perfectly matches the ground-truth communities. (See Fig.\ref{DRAW_KARATE} and Fig.\ref{DRAW_Polbooks})

\item In terms of \emph{GK-Means}, we set the susceptibility parameter sufficiently large (0.9) on \emph{Karate} so that agents’ opinions within the same community converge to consensus, but the divided communities is significantly different from the ground-truth (see \ref{de:Initial_consensus}); we set a smaller susceptibility parameter (0.4) on \emph{PolBooks} to make the divided communities as close to the ground-truth as possible (see Fig.\ref{DRAW_Polbooks}), but this leads to significant differences in opinions within the same community (small \emph{ACL});

\item For \emph{LPA-HK}, as the confidence level $\delta$ increases, the evolution of group opinions will move from pluralism to polarization and eventually consensus \cite{HK}. We set $\delta = 0.95$ on both on \emph{Karate} and \emph{PolBooks}, and all agents will converge to consensus. But again, the community divisions obtained are very low-quality (see \ref{de:Initial_consensus}).

\end{enumerate}

Hence, both \emph{GK-Means} and \emph{LPA-HK} algorithms have the ability to foster agreement among agents belonging to the same community, but at the cost of generating community divisions of subpar quality. The \emph{GCAOFP} model effectively combines the processes of opinion evolution and community building in order to attain the localized convergence of opinions. Within the context of \textbf{community dynamics process}, \emph{GCAOFP} establishes a correlation between the optimization target, which is social welfare, and the extent of local consensus that is achieved. As per Eq.\ref{Y_define},

\begin{equation}
	\begin{aligned}
		\label{Community_CDP}
    \mathbf{Y}_{ij}\left ( T \right ) &= \left ( \lambda_{i} \mathbf{D}_{ij}\left ( T \right ) - \mathbf{\mathbf{B}}_{ij}\left ( T \right ) \right ) \mathbf{W}_{ij}+ \left ( \lambda_{j}\mathbf{D}_{ji}\left ( T \right ) - \mathbf{\mathbf{B}}_{ji}\left ( T \right ) \right ) \mathbf{W}_{ji} \\
    &= \left ( \lambda_{i} \mathbf{W}_{ij} + \lambda_{j} \mathbf{W}_{ji} \right ) \mathbf{D}_{ij}\left ( T \right ) - \left ( \mathbf{W}_{ij} + \mathbf{W}_{ji}\right ) ReLu\left ( \mathbf{D}_{ij}\left ( T \right ) \right )\\
    &= \left\{\begin{matrix}
  & \left [ \left ( \lambda_{i}- 1 \right ) \mathbf{W}_{ij} +  \left ( \lambda_{j}- 1 \right ) \mathbf{W}_{ji} \right ] \mathbf{D}_{ij}\left ( T \right ) , \enspace \mathbf{D}_{ij}\left ( T \right ) > 0  \\
  & \left ( \lambda_{i} \mathbf{W}_{ij} + \lambda_{j} \mathbf{W}_{ji} \right ) \mathbf{D}_{ij}\left ( T \right ), \enspace \mathbf{D}_{ij}\left ( T \right ) \le 0
\end{matrix}\right.
   \end{aligned}
\end{equation}

\noindent Therefore, $\mathbf{Y}_{ij}\left ( T \right )$ shows a positive linear correlation of $\mathbf{D}_{ij}\left ( T \right )$. During each iteration, each agent adopts the community label that maximizes her utility (Eq.~(\ref{modify_utility})), thus the consensus level of the opinions of agents within the same community will also increase as the community structure is updated. In the \textbf{opinion formation process}, \emph{GCAOFP} assigns greater weight to the opinions of intra-community neighbors, which further promotes the local convergence of opinions. These two operations guarantee that \emph{GCAOFP} continuously increases the level of consensus among members within the identified communities during the repeated iterations given any initial settings.

\subsubsection{Scalability Analysis.}

We compare \emph{GCAOFP} with six baseline tools on the scalability, specifically, for fair comparison, all experiments were performed on the same machine (with Intel(R) Xeon(R) Gold 6326 CUP (2.90 GHz), 320 GB of RAM), and in order to eliminate the effect of randomness, all the results were averaged by multiple implementations using 100 different initial settings on the same network. Fig.\ref{fig:RUN_TIMES} shows the execution time of different dynamics models under different scales of link numbers, from which we can observe that \circled{1} the four standard opinion dynamics models are significantly faster than the community-related models, and this advantage becomes more pronounced as $M$ increases; \circled{2} \emph{GCAOFP} is the fastest among the three community-related models, especially on the first three smaller networks, where the running time of \emph{GCAOFP} was reduced by more than two orders of magnitude on average compared to \emph{GK-Means} and \emph{LPA-HK}. The overall time complexity of \emph{LPA-HK} is $\mathcal{O} \left ( n^{2}  \right )$, and \emph{GK-Means} is $\mathcal{O} \left ( t^{Exp} M \left ( \left \langle K \right \rangle + 1 \right ) \right )$, where $\left \langle K \right \rangle = \frac{M}{n}$ denotes the average degree of agents. The \emph{best-response dynamics} requires fewer iterations (smaller $t^{Exp}$) on average to find the optimal community structure when $n$ is smaller, therefore, on scalability, \emph{GCAOFP} is better than \emph{GK-Means} and \emph{LPA-HK}, especially on smaller plain networks.

\subsubsection{Parameter Analysis.}

Higher $\lambda$ implies higher susceptibility of the agents to their community labels, the faster the opinions of agents within the same community converge to consensus, and \emph{GCAOFP} is more likely to divide more communities; $\psi$ works similarly to the confidence threshold $\varrho$ (Eq.\ref{HK_DEFINE}) defined in \emph{HK}, where a larger $\psi$ could better promote the opinion exchange among agents (agents can accept more opinions of neighbors), which will further promote the group opinions to reach consensus, and also indicating that \emph{GCAOFP} is more likely to divide a smaller number of communities. 

\begin{figure*}
	\centering
    \begin{minipage}{0.99\linewidth}
		\includegraphics[width=1\textwidth]{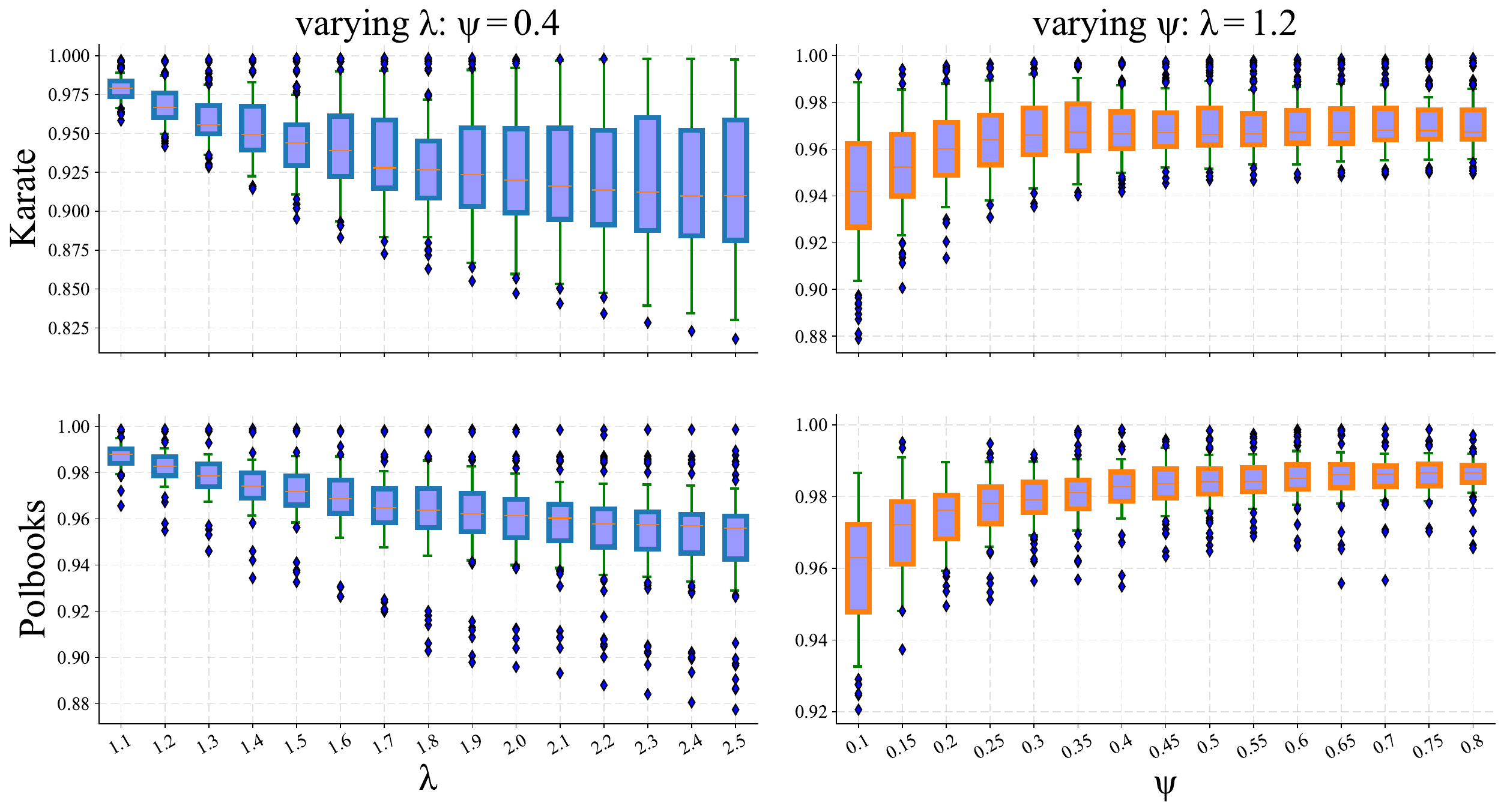}
        \caption{The boxplots of \emph{RCSW} (100 implements) of \emph{GCAOFP} on \emph{Karate} varying $\lambda$ and $\psi$, respectively.}
        \label{fig:parameters}
    \end{minipage}

    \qquad
    
	\begin{minipage}{0.99\linewidth}
		\centering
        \includegraphics[width=2.42in]{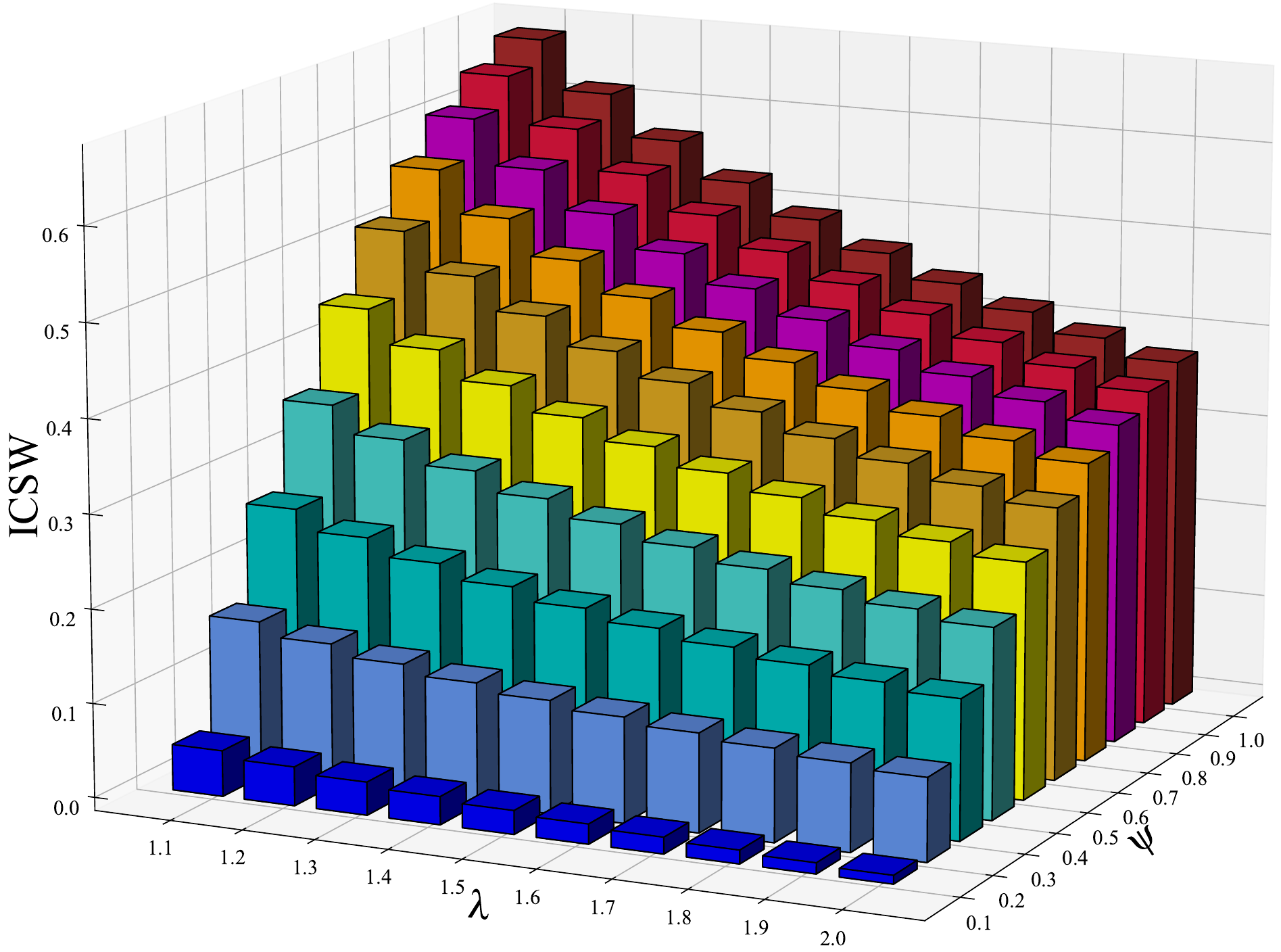}
        \hspace{0.03in}
	  \includegraphics[width=2.5in]{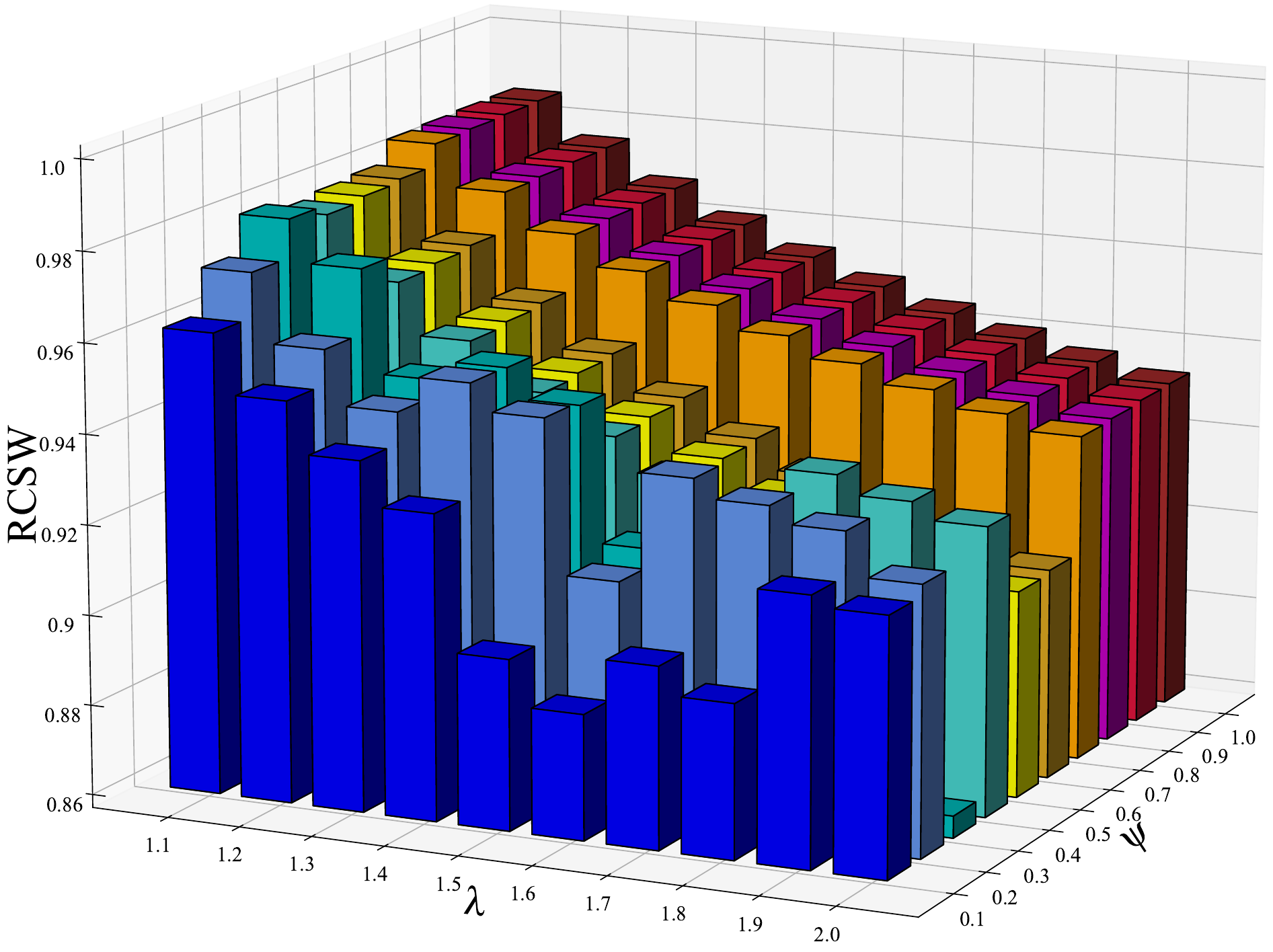}
        \caption{Given a certain initial opinion vector and community member matrix, the changes of \emph{ICSW} and \emph{RCSW} of \emph{GCAOFP} on \emph{Karate} under different $\lambda$ and $\psi$.}
        \label{fig:certain_opinion_community}
	\end{minipage}
	
\end{figure*}

We have selected two smaller networks, namely \emph{Karate} and \emph{Polbooks}, as case studies to examine the influence of $\lambda$ and $\psi$. We will manipulate the parameters $\psi$ and $\lambda$ to observe their relative effects. Fig.\ref{fig:parameters} displays the boxplots of \emph{RCSW} acquired by the \emph{GCAOFP} method. The data includes numerous implementations, each with 100 distinct beginning opinion vectors and community membership matrices. These implementations were conducted using various parameter values for $\lambda$ and $\psi$. It is evident that: Under the assumption that $\psi$ remains constant, \emph{GCAOFP} is capable of achieving a higher \emph{RCSW} with reduced variance when the parameter $\lambda$ is assigned a smaller value. This implies that \emph{GCAOFP} has a preference for dividing a smaller number of communities, leading to a convergence of agents' opinions on a global scale. As the parameter $\lambda$ increases, \emph{GCAOFP} exhibits a decrease in the \emph{RCSW} with a corresponding rise in variance. This trend suggests that \emph{GCAOFP} is capable of generating a wider range of community division outcomes, leading to a progressive shift in opinions from global convergence to local convergence. Hence, it can be inferred that the augmentation of $\lambda$ fosters the local convergence of beliefs while simultaneously encouraging the fracturing of community. When the value of $\psi$ is increased while $\lambda$ remains constant, the resulting \emph{RCSW} value will be bigger and the variance will be reduced. This indicates that the distribution of opinions becomes more concentrated, providing evidence that increasing $\psi$ can facilitate the global convergence of beliefs and mitigate community division. Intuitively, when the value of the community susceptibility parameter $\lambda$ is set to be sufficiently enough, the \emph{GCAOFP} algorithm is able to strike a more favorable balance between the local convergence of opinions and the attainment of high-quality community divides by changing the opinion threshold $\psi$.

Finally, we use \emph{Karate} as an example to show the effect of setting different $\lambda$ and $\psi$ on the level of consensus reached among agents under a given initial opinion vector and community member matrix. Fig.\ref{fig:certain_opinion_community} illustrates the experimental results, from which we can observe: \circled{1} Keep $\lambda$ unchanged, \emph{RCSW} is sensitive to changes in $\psi$ when $\psi \le 0.4$, and the value of \emph{RCSW} will remain unchanged when $\psi \ge 0.7$; \circled{2} When $\psi \ge 0.4$, increasing the value of $\lambda$ will lead to a continuous decrease in the value of \emph{RCSW}, which, combined with the definition of \emph{RCSW}, means that \emph{GCAOFP} allows for a greater diversity of opinions. When $\psi \le 0.4$, this pattern will still exist within the intervals ($\lambda \in \left [ 0.1, 0.3 \right ]$, for example); \circled{3} Both fixed $\lambda$ increasing $\psi$ and fixed $\psi$ increasing $\lambda$ can increase the value of \emph{ICSW}. This is reasonable because: 1) when the community structure is unchanged, increasing $\psi$ can reduce the occurrence of situations where the opinions of intra-community neighbors outside the confidence level, and on the other hand, it can increase the inter-community utility; 2) reducing $\lambda$ can mitigate the adverse effects of penalties due to the opinions of intra-community neighbors outside the confidence level when agents' opinions are fixed. In addition, as shown in the figure, the former has a greater impact on social welfare.

\subsection{Community detection}

In order to validate the effectiveness of \emph{GCAOFP} on the community detection task, we proceed to conduct case studies on four benchmarks. Figure 8 displays the boxplots of \textbf{A}djusted \textbf{R}and \textbf{I}ndex (\emph{ARI}) \cite{ARI} and \textbf{A}djusted \textbf{M}utual \textbf{I}nformation (\emph{AMI}) \cite{AMI} acquired using various algorithms, namely \emph{LPA} \cite{LPA}, \emph{k-clique} \cite{k-clique}, \emph{GLEAM} \cite{GLEAM}, \emph{GK-Means}, \emph{LPA-HK}, and \emph{GCAOFP} (referred to as \emph{GCAOFP-O}), on real-world networks. These algorithms were executed using 100 distinct initial opinion vectors, with the ground-truth communities provided. It is evident that \emph{GK-Means}, \emph{LPA-HK}, and \emph{GCAOFP} models demonstrate notable variations in performance when subjected to varied initial opinions. Conversely, the performance of the remaining models does not exhibit substantial fluctuations. Additionally, we conducted performance measurements on \emph{GCAOFP} using a predetermined initial opinion vector and 100 distinct initial community structures, denoted as \emph{GCAOFP-C}. Multiple simulation results indicate that: \circled{1} Across all networks, opinion-related models exhibit comparable or even superior performance compared to models that solely rely on network structures; \circled{2} As the network size increases, the variability in the performance of all opinion-related models tends to decrease, implying a diminishing significance of individual opinions; \circled{3} \emph{GCAOFP}, in comparison to the initial community structure, displays higher sensitivity to the initial opinions of the agents. In other words, unless in exceptional cases, \emph{GCAOFP} can eventually yield the same stable community structure given the same opinion vector; \circled{4} When provided with a reasonable opinion vector, \emph{GCAOFP} has the capability to identify communities that are more closely aligned with the actual community structure compared to other models.

\begin{figure}[htbp]
	\centering

    \begin{minipage}{0.99\linewidth}

    \centering
    \linespread{0.1}
    \includegraphics[width=1\textwidth]{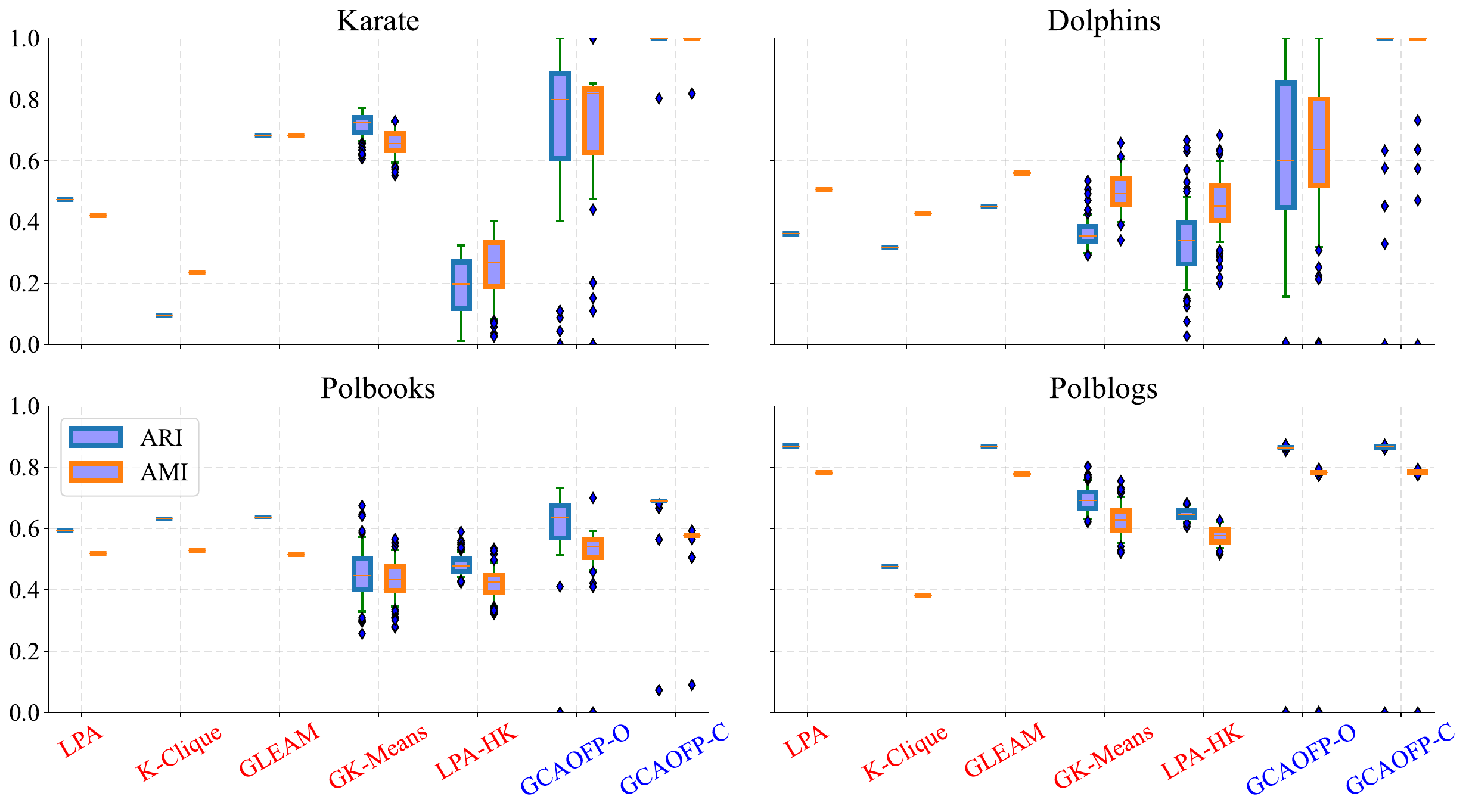}
    \caption{The illustration of \emph{LPA}, \emph{K-Clique}, \emph{GLEAM}, \emph{GK-Means}, \emph{LPA-HK} and \emph{GCAOFP} on four real-world networks with ground-truth communities under \emph{ARI} and \emph{AMI} (100 implements).}
    \label{fig:community_detection}
    
    \end{minipage}
 
\end{figure}

\begin{figure}[htbp]
	\centering

	\begin{minipage}{0.99\linewidth}
		\centering
		\subfloat[\emph{GK-Means}]{
        \includegraphics[width=2.1in]{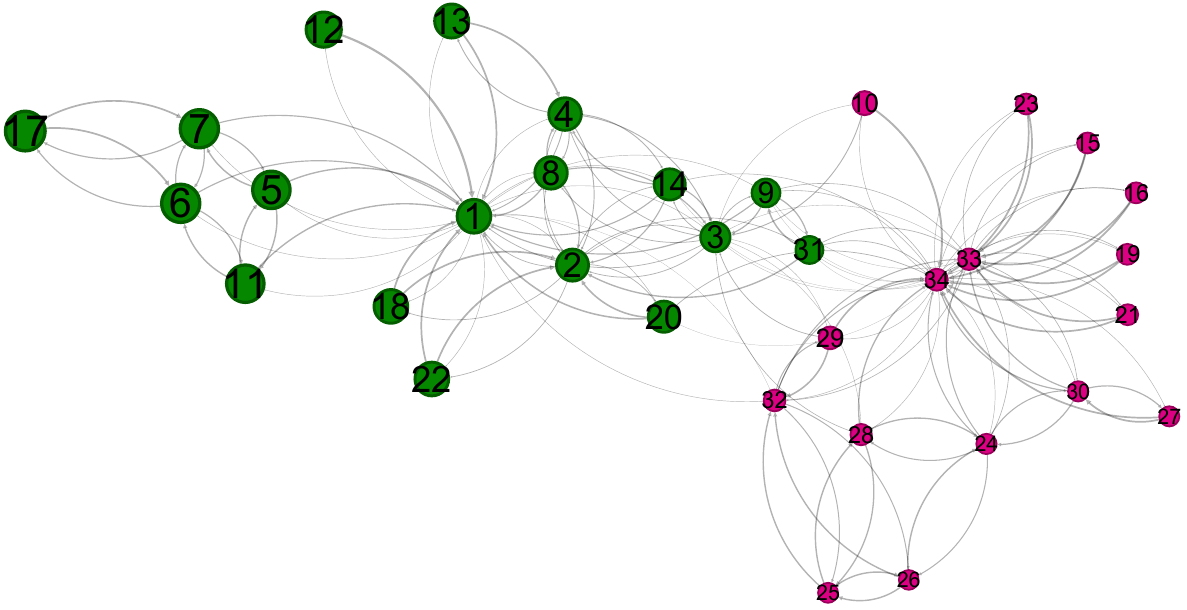}
        }
        \subfloat[\emph{LPA-HK}]{
	   \includegraphics[width=2.1in]{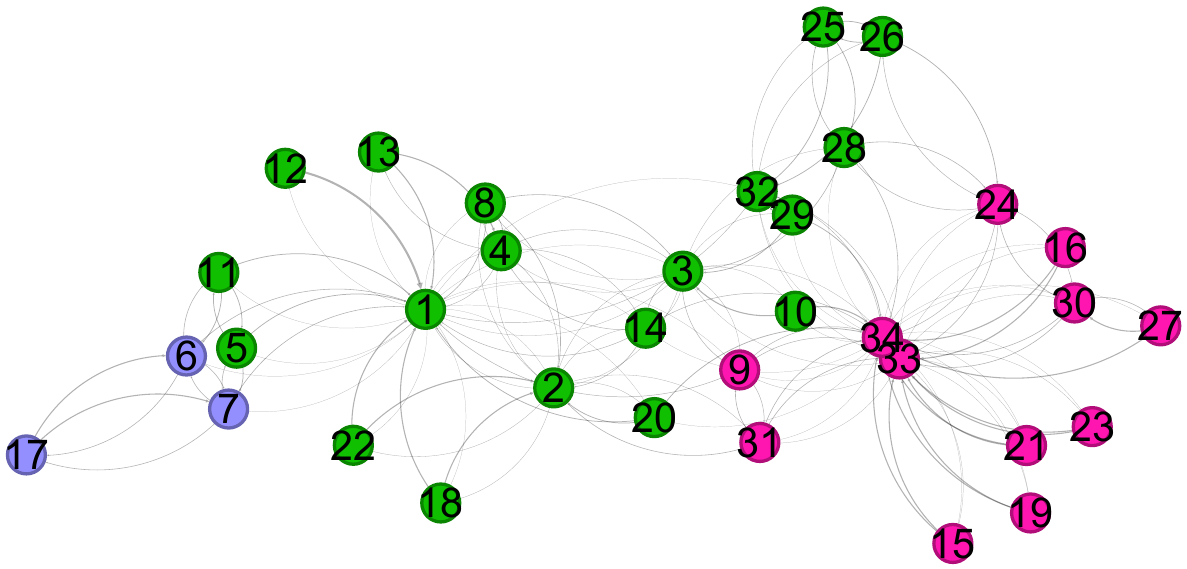}
        }
        \quad    
        \subfloat[\emph{GCAOFP}]{
    	   \includegraphics[width=2.1in]{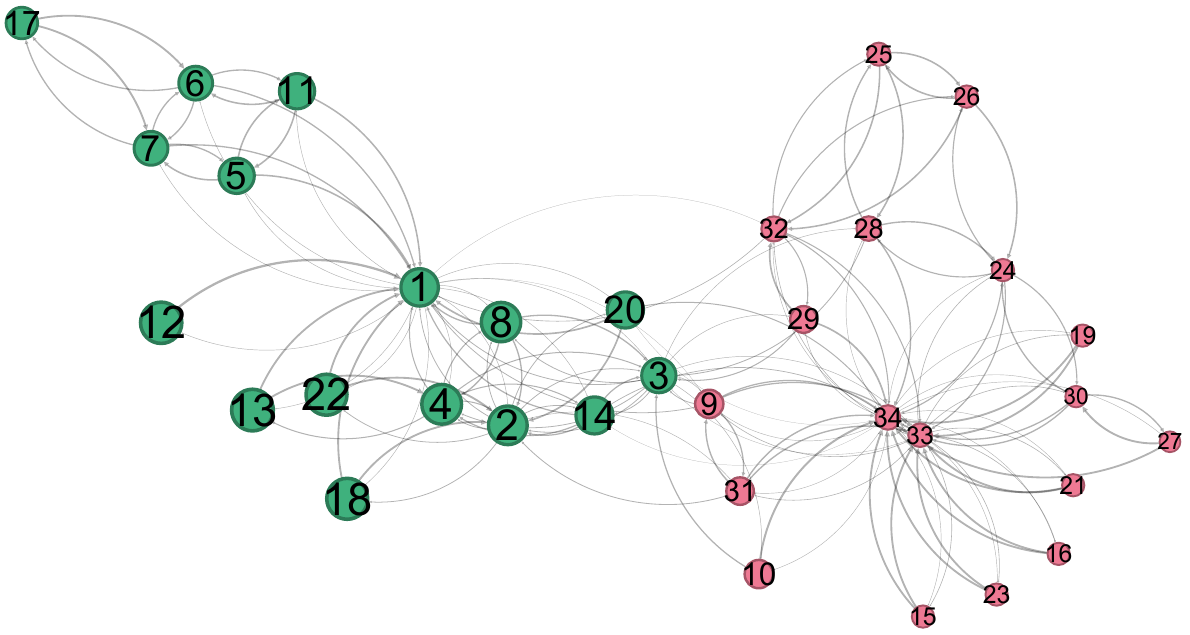}
        }
        \subfloat[Ground-Truth]{
	   \includegraphics[width=2.1in]{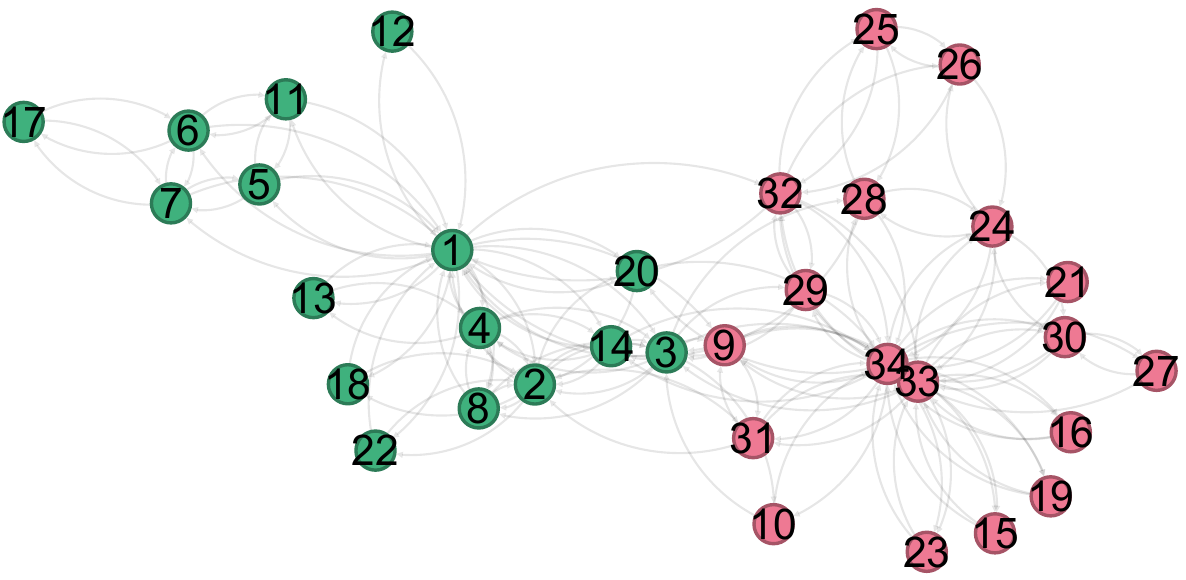}
        }
        \caption{The communities detected by different dynamics models on \emph{Karate}.}
        \label{DRAW_KARATE}
	\end{minipage}
\end{figure}

\begin{figure}[htbp]
	\centering
	\begin{minipage}{0.99\linewidth}
		\centering
		\subfloat[\emph{GK-Means}]{
            \includegraphics[width=2.1in]{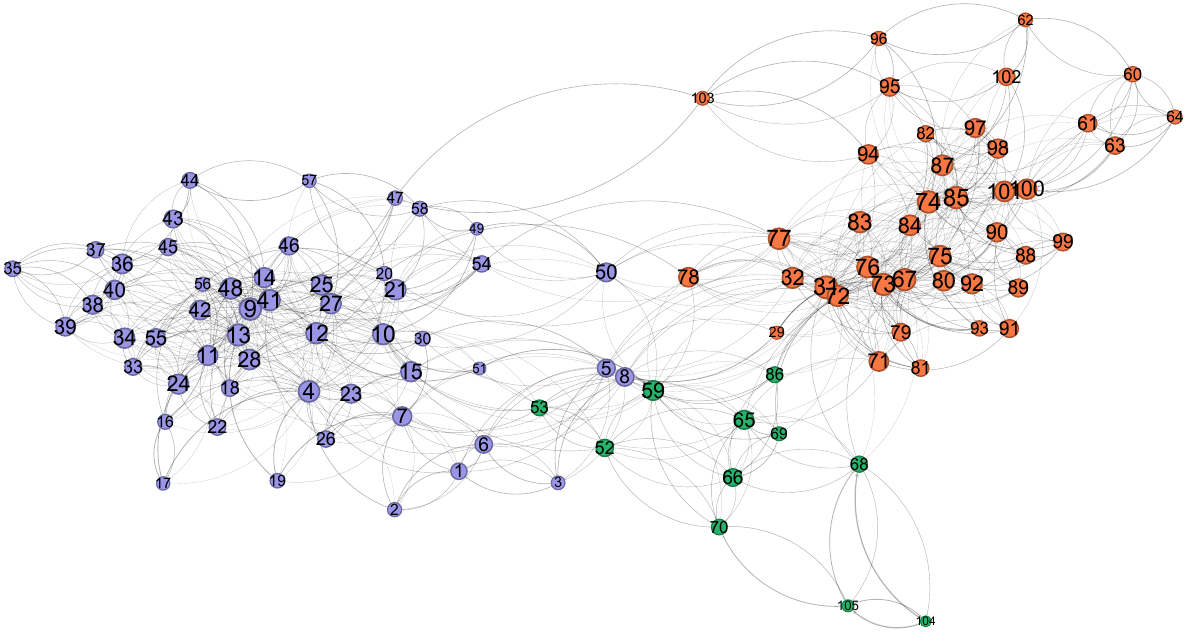}
        }
        \subfloat[\emph{LPA-HK}]{
	   \includegraphics[width=2.1in]{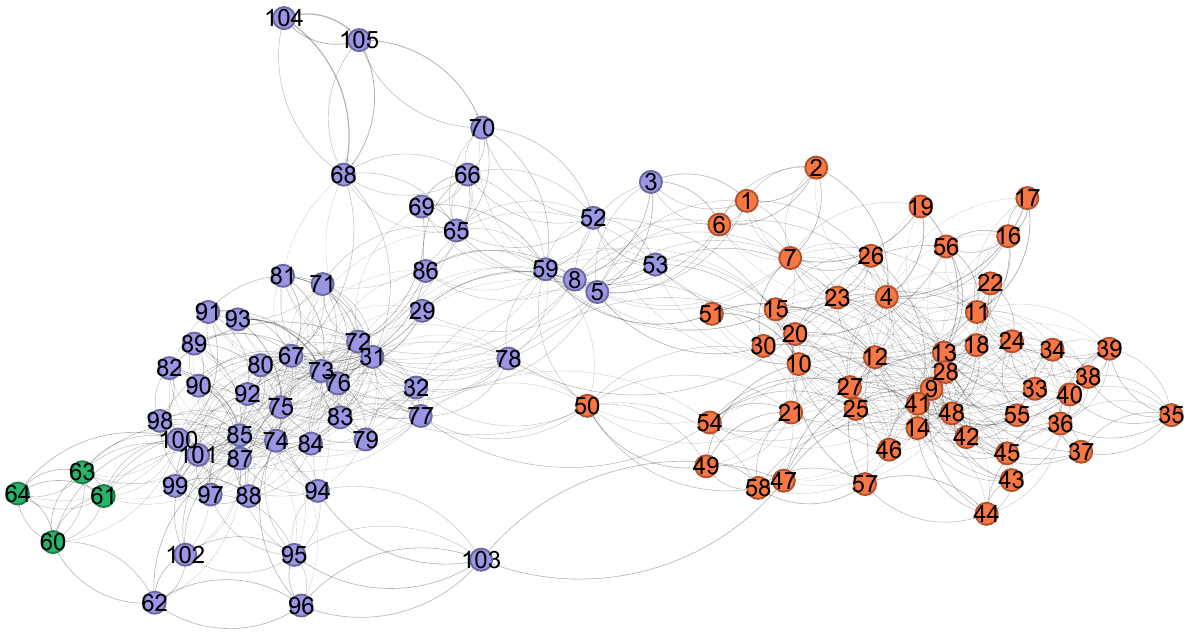}
        }
        \quad    
        \subfloat[\emph{GCAOFP}]{
    	   \includegraphics[width=2.1in]{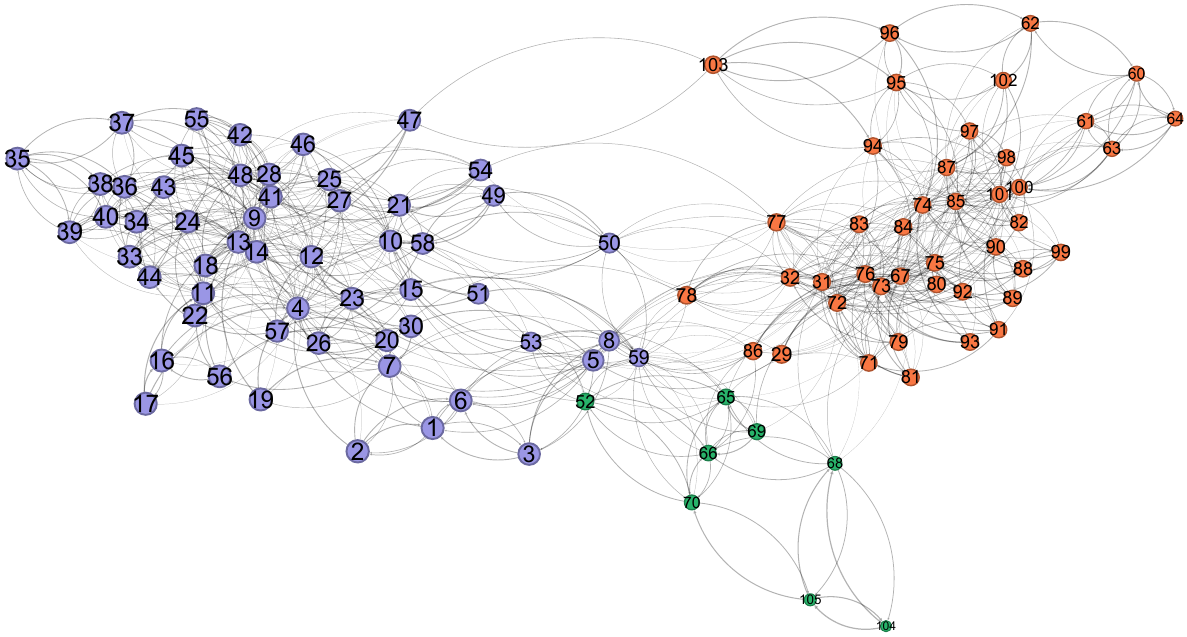}
        }
        \subfloat[Ground-Truth]{
	   \includegraphics[width=2.1in]{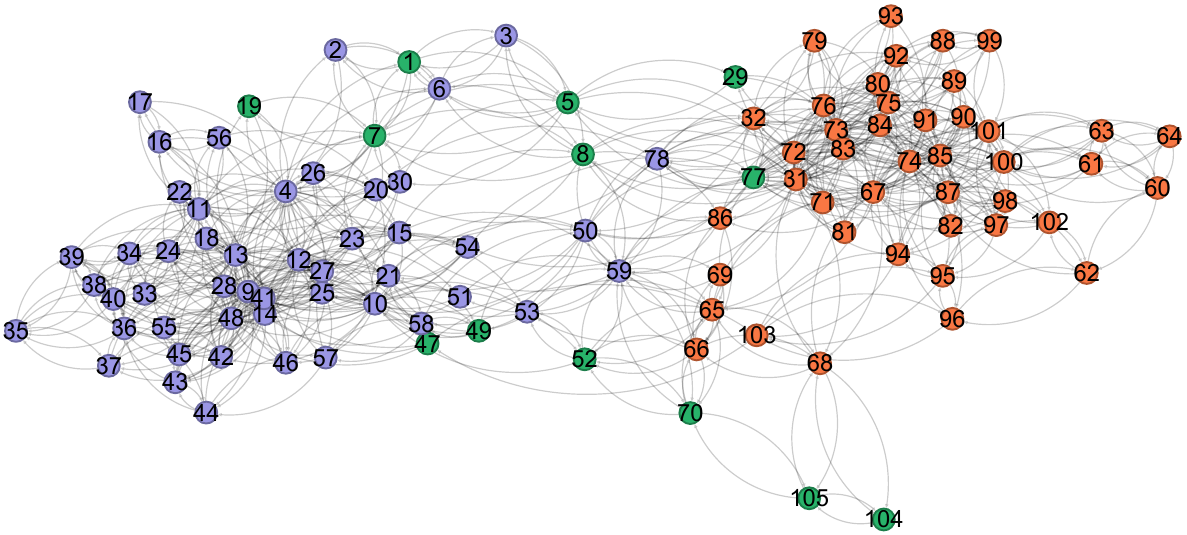}
        }
        \caption{The communities detected by different dynamics models on \emph{PolBooks}.}
        \label{DRAW_Polbooks}
    \end{minipage}
\end{figure}

In order to illustrate the most accurate outcomes of community discovery using various dynamics models, we have chosen two smaller networks, \emph{Karate} and \emph{PolBooks}. These networks are visualized in Fig.\ref{DRAW_KARATE} and Fig.\ref{DRAW_Polbooks}, respectively. In all network snapshots, with the exception of the ground-truth snapshot, the size of nodes corresponds to their final opinions. The color of nodes is determined by their community labels, while the thickness of links between nodes indicates the influence of the target node on the source node. The hue of the nodes represents the actual community labels in the ground-truth snapshots. The results presented in Fig.\ref{DRAW_KARATE} demonstrate that the communities identified by \emph{GCAOFP} exhibit a high degree of accuracy in matching the ground-truth communities. Specifically, all nodes are correctly partitioned into their respective communities, as evidenced by the \emph{ARI} and \emph{AMI} scores of 1. In contrast, both \emph{GK-Means} and \emph{LPA-HK} algorithms exhibit a higher rate of misclassification, as indicated by their lower \emph{ARI} and \emph{AMI} scores. For \emph{GK-Means}, the \emph{ARI} and \emph{AMI} scores are 0.7717 and 0.7268, respectively, while for \emph{LPA-HK}, the \emph{ARI} and \emph{AMI} scores are 0.3236 and 0.4032, respectively.

Upon examination of Fig.\ref{DRAW_Polbooks}, it is evident that \emph{GCAOFP} partitions a collective of 12 nodes into an incorrect community, resulting in an \emph{ARI} of 0.7270 and \emph{AMI} of 0.6366. In contrast, \emph{GK-Means} and \emph{LPA-HK} misclassify 16 and 21 nodes, respectively. Specifically, \emph{GK-Means} achieves an \emph{ARI} of 0.6745 and \emph{AMI} of 0.5652, while \emph{LPA-HK} yields an \emph{ARI} of 0.5896 and \emph{AMI} of 0.5280. The observation can be made that the effectiveness of community partitioning for all models on the \emph{PolBooks} dataset is comparatively lower than that on the \emph{Karate} dataset. This discrepancy can be attributed to the sparsity of connections among agents inside the green communities in \emph{PolBooks}, which therefore diminishes the impact of intra-community neighbors. In the case of \emph{GCAOFP} and \emph{GK-Means} algorithms, it was observed that all nodes belonging to the green community, with the exception of nodes 52, 70, 104, and 105, were inaccurately partitioned. Furthermore, the \emph{LPA-HK} algorithm exhibited a division of all nodes inside the green community into two separate communities.

\begin{wrapfigure}{l}{7cm}
\center
\includegraphics[width=0.5\textwidth]{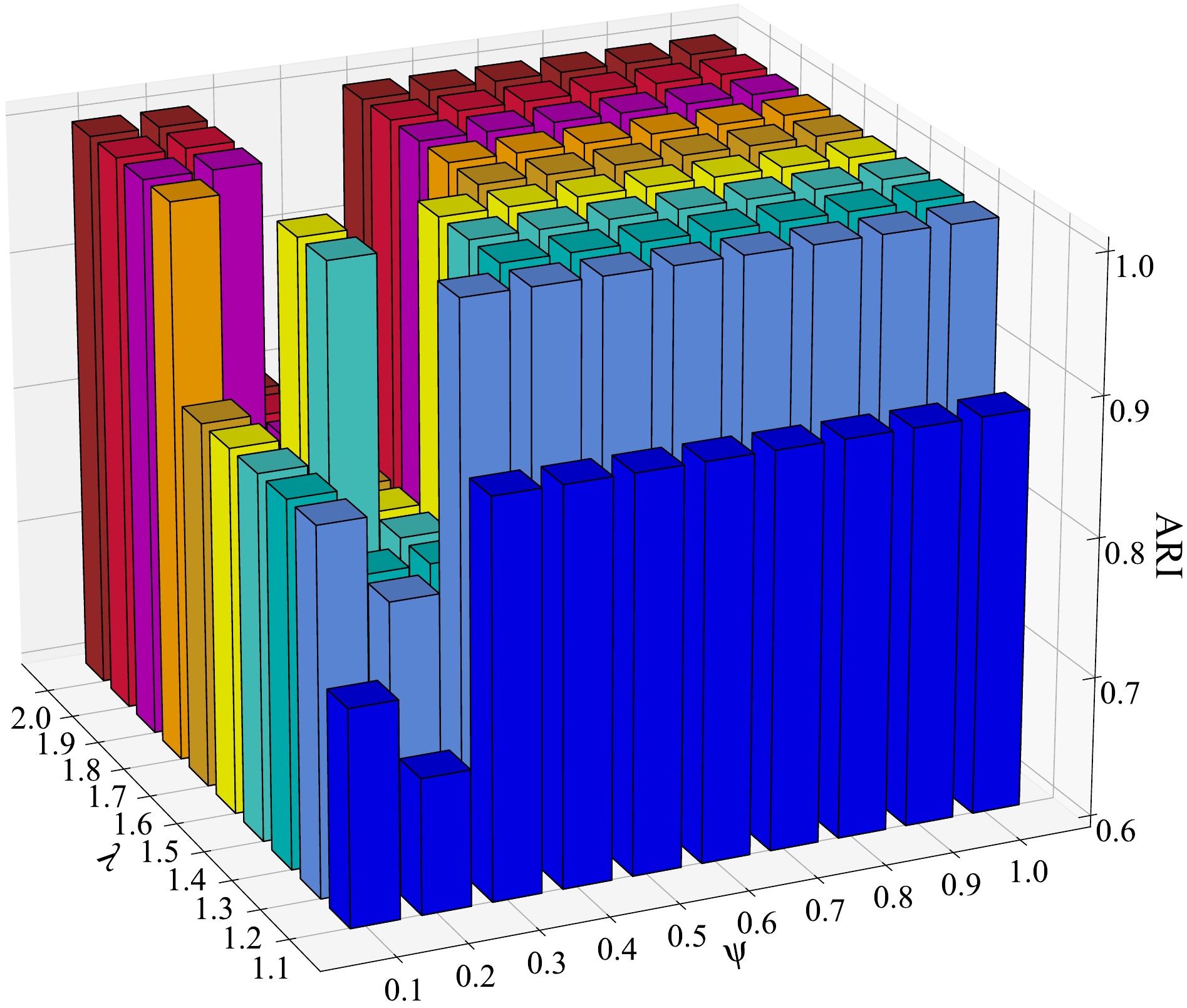}
\caption{Given initial opinions and community member matrix, \emph{ARI} under different $\lambda$ and $\psi$.}
\label{fig:ARI_AMI_CERTAIN}
\end{wrapfigure}

In conclusion, we shall now analyze the influence of parameters $\lambda$ and $\psi$ on the overall quality of the segregated communities. Fig.\ref{fig:ARI_AMI_CERTAIN} displays the \emph{ARI} achieved by \emph{GCAOFP} on the \emph{Karate} network. The \emph{ARI} is evaluated for various values of $\lambda$ and $\psi$, while keeping the initial opinion vector and community member matrix constant. The results indicate that superior partitions are attainable when the confidence level ($\psi$) and community sensitivity ($\lambda$) exceed the critical values of 1.2 and 0.5, respectively. \ref{de:Initial_state} illustrates the initial state, revealing a significant divergence of beliefs across adjacent agents. Consequently, increasing the value of $\psi$ will promote the convergence of attitudes within communities and lead to a more effective division of higher quality communities.

\begin{figure}[htbp]
\center
\linespread{0.1}
\includegraphics[width=1\textwidth]{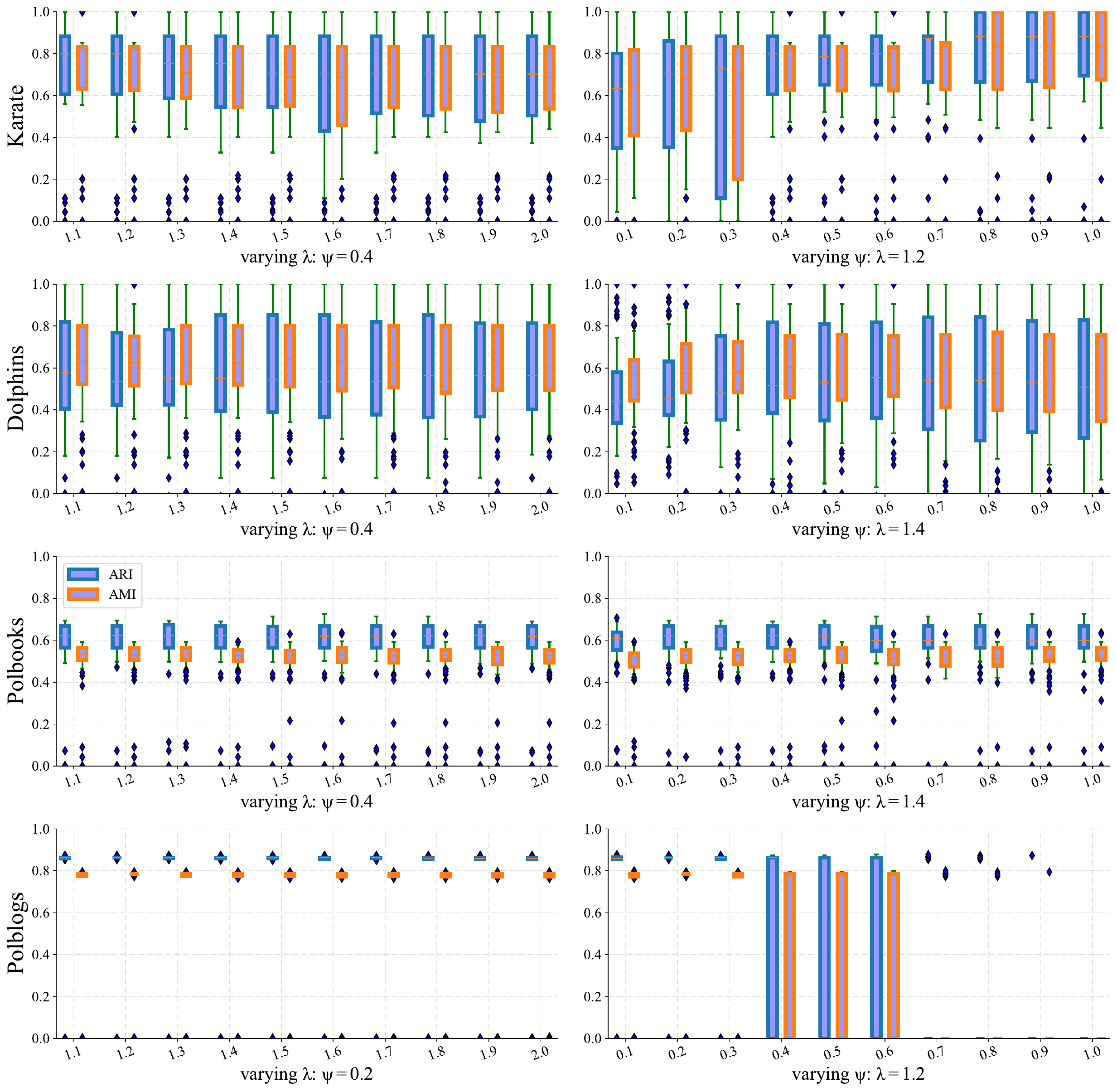}
\caption{The boxplots of \emph{ARI} and \emph{AMI} (100 implements) of \emph{GCAOFP} on real-world networks.}
\label{fig:community_parameter}
\end{figure}

Fig.\ref{fig:community_parameter} illustrates the variations in \emph{ARI} and \emph{AMI} across different values of $\lambda$ and $\psi$ during the process of community recognition on four actual networks, using a total of 100 distinct initial states. It is evident that altering various factors has the potential to influence the outcomes of community partitioning in \emph{GCAOFP}. One notable observation is that inside the \emph{POLBLOGS} network, when the value of $\psi$ is less than 0.4, the communities identified by the \emph{GCAOFP} algorithm exhibit a high degree of accuracy in aligning with the ground-truth communities. Nevertheless, as the variable $\psi$ progressively increases, the \emph{GCAOFP} algorithm tends to divide the whole network into a single community, leading to unfavorable partitioning outcomes. In the majority of initial states, where the value of $\psi$ is equal to or greater than 0.7, the network will ultimately be partitioned into a single community. Consequently, both the \emph{ARI} and \emph{AMI} will converge to a value of 0.

\section{Conclusion}

This study aims to examine the interconnected dynamics of community formation and opinion evolution within real-world social systems. Additionally, we introduce a novel model for opinion dynamics, namely \emph{GCAOFP}, which incorporates bounded confidence. The method of updating opinions, as proposed by \emph{GCAOFP}, considers the community structure among agents, which is typically overlooked in the conventional opinion dynamics model. \emph{GCAOFP} introduces two novel opinion update rules. 1) Agents tend to place greater faith in the ideas expressed by those inside their own community. 2) Agents have a higher propensity to accept opinions that align closely with their own beliefs, while selectively disregarding opinions that fall beyond their established confidence threshold. To enhance compliance with rule 1, we propose the introduction of a community formation game that takes into account the opinion vector. In this game, each agent's utility function is associated with its respective community label. Each individual agent employs reaction dynamics in order to optimize their usefulness, while asynchronously updating their community label. The aforementioned update process is iteratively performed until all agents reach a state where further improvement in their utilities is not possible. Ultimately, the process yields the optimal community partition that maximizes social welfare. The convergence of the co-evolutionary dynamics process to an equilibrium state, characterized by a steady opinion vector and community member matrix, is theoretically demonstrated for any initial settings within a finite number of iterations. A number of studies were conducted to evaluate the effectiveness and efficiency of \emph{GCAOFP}, as well as to ascertain the significance of two specific characteristics. In this study, the \emph{GCAOFP} model was employed to address the community detection problem. The model's performance was evaluated by comparing it to five baseline methods using two evaluation metrics, namely \emph{ARI} and \emph{AMI}. In summary, the \emph{GCAOFP} approach offers a novel research framework for investigating the evolutionary dynamics of opinions inside networked multi-agent systems. Furthermore, we posit that the integration of agent opinions with network structure might be used to additional tasks, such as link prediction.

\backmatter





\bmhead{Acknowledgements}

This research was partially supported by National Natural Science Foundation of China under Grants 72371133, 71871109, 62076060, 71801123 and 61932007, in part by National Social Science Funds of China under Grant 22BGL261, in part by General Project of Philosophy and Social Science Research in Jiangsu Universities under Grant 2021SJA0337, and in part by the Funding for Research on Philosophy and Social Science in Jiangsu Province under Grant 2023SJYB0181.

\bmhead{Author Contributions}

Conceptualization: SFZ; Formal Analysis: SFZ; Writing—original draft: SFZ; Software: SFZ; Funding acquisition: ZB; Investigation: ZB; Methodology: SFZ; Writing—review \& editing: XTS; Project administration: ZB; Supervision: ZB; Validation: XTS; Data curation: XTS; Resources: ZB; Visualization: SFZ.

\section*{Declarations}

\bmhead{Conflict of interest} 

The authors have no conflicts of financial or proprietary interests in any material discussed in this paper.

\bmhead{Code availability}

The codes of this work are available at \href{https://github.com/ZINUX1998/GCAOFP}{\texttt{https://github.com/ZINUX1998/GCAOFP}}.









\clearpage

\begin{appendices}

\section{Supporting Information}





\subsection{Feature of \emph{DGSWG}}\label{de:Feature_DGSWG}

\noindent\textsc{Proof:} Under the existing community structure $\mathbf{S}= \left [ \mathbf{S}_{1},\mathbf{S}_{2},\dots ,\mathbf{S}_{n} \right ]$, if any agent $i$ changes her community label from $\mathbf{S}_{i}$ to $\mathbf{S}_{i}^{\prime }$, the accompanying utility change is

\begin{equation}
	\begin{aligned}
		\label{ao_2}
    \Delta u_{i}^{\mathbf{S}_{i} \to \mathbf{S}_{i}^{\prime } }= u_{i}^{\ast }\left ( \left [ \mathbf{S}_{i}^{\prime },\mathbf{S}_{-i} \right ]  \right ) -u_{i}^{\ast }\left ( \left [ \mathbf{S}_{i},\mathbf{S}_{-i} \right ]  \right ) =\sum_{j}\left ( \mathbf{S}_{i}^{\prime }-\mathbf{S}_{i} \right )\mathbf{S}_{j}^{\mathsf{T}}\mathbf{Y}_{ij}
   \end{aligned}
\end{equation}

the potential value change is

\begin{equation}
	\begin{aligned}
		\label{ao_2}
    \Delta \phi ^{\mathbf{S}_{i} \to \mathbf{S}_{i}^{\prime } } &= \phi^{\ast}\left ( \left [ \mathbf{S}_{i}^{\prime },\mathbf{S}_{-i} \right ] \right ) -\phi^{\ast}\left ( \left [ \mathbf{S}_{i},\mathbf{S}_{-i} \right ] \right ) \\
    &= \frac{1}{2}\Delta u_{i}^{\mathbf{S}_{i} \to \mathbf{S}_{i}^{\prime } } + \frac{1}{2}\sum_{j\ne i}\left [ \left ( \mathbf{S}_{j}\left ( \mathbf{S}^{\prime } \right )_{i}^{\mathsf{T}}  \mathbf{Y}_{ij} +\sum_{k\ne i}\mathbf{S}_{j}\mathbf{S}_{i}^{\mathsf{T}}Y_{jk} \right )- \left ( \mathbf{S}_{j}\mathbf{S}_{i}^{\mathsf{T}}\mathbf{Y}_{ij} +\sum_{k\ne i}\mathbf{S}_{j}\mathbf{S}_{i}^{\mathsf{T}}Y_{jk} \right )\right ] \\
    &= \frac{1}{2}\Delta u_{i}^{\mathbf{S}_{i} \to \mathbf{S}_{i}^{\prime } } + \frac{1}{2}\sum_{j\ne i}\left [ \mathbf{S}_{j}\left ( \mathbf{S}^{\prime } \right )_{i}^{\mathsf{T}}  \mathbf{Y}_{ij} - \mathbf{S}_{j}\mathbf{S}_{i}^{\mathsf{T}}\mathbf{Y}_{ij}\right ] \\
    &= \frac{1}{2}\Delta u_{i}^{\mathbf{S}_{i} \to \mathbf{S}_{i}^{\prime } } + \frac{1}{2}\sum_{j\ne i}\left ( \mathbf{S}_{i}^{\prime }-\mathbf{S}_{i} \right )\mathbf{S}_{j}^{\mathsf{T}}\mathbf{Y}_{ij}\\
    &= \Delta u_{i}^{\mathbf{S}_{i} \to \mathbf{S}_{i}^{\prime } }
   \end{aligned}
\end{equation}

Therefore, \textbf{\emph{DGSWM}} is a potential game with a finite size of strategies, and there exists at least one pure Nash equilibrium.

\subsection{Effectiveness of \emph{CDP}}\label{de:Effectiveness_CDP}

\noindent\textsc{Proof:} The \emph{best-response dynamics} will eventually halt after $\tau \in \mathbb{R} ^{+}$ iterations because the game is finite, accompanied by a sequence of community member matrices $\mathbf{S}\left ( 0 \right ), \mathbf{S}\left ( 0 \right ), \dots, \mathbf{S}\left ( \tau \right ) \left ( \mathbf{S}\left ( 0 \right ) = \mathbf{s}\left ( T \right ) , \mathbf{S}\left ( \tau \right ) = \mathbf{s}\left ( T + 1 \right ) \right )$. During each iteration, each agent adopts the community label that maximizes her utility (Eq.~(\ref{modify_utility})) in an asynchronous manner, thus $\forall \xi, 1 \le \xi  \le \tau $, we have $\phi ^{\ast }  \left ( \mathbf{S}\left ( \xi \right ) \right ) \ge \phi ^{\ast }  \left ( \mathbf{S}\left ( \xi - 1 \right ) \right )$, therefore $tr\left \{\left ( \mathbf{S}\left ( \tau \right ) \right ) ^{\mathsf{T}} \left [ \left ( \Lambda \mathbf{D}\left ( T \right ) - \mathbf{B}\left ( T \right ) \right ) \odot \mathbf{W} \right ] \mathbf{S}\left ( \tau \right ) \right \} \ge tr\left \{\mathbf{S}^{\mathsf{T}}\left ( t \right )\left [ \left ( \Lambda \mathbf{D}\left ( T \right ) - \mathbf{B}\left ( T \right ) \right ) \odot \mathbf{W} \right ] \mathbf{S}\left ( t \right ) \right \}, t = 0,1,\cdots,\tau - 1$, i.e., $\mathbf{S}\left ( \tau \right )$ is the solution of \textbf{\emph{CDSWM}}. Given that $\mathbf{x} \left ( T \right )$, $\mathbf{W}$ and $\Lambda $ remain unchanged, the social welfare (Eq.~(\ref{Social_welfare})) associated with \textbf{\emph{CDP}} will be continuously optimized. According to the termination condition of \emph{best-response dynamics}, in the context of $\mathbf{S}\left ( \tau \right )$, all agents are unable to improve their utility by changing their community labels, so $\mathbf{S}\left ( \tau \right )$ will be a pure Nash equilibrium w.r.t. \textbf{\emph{DGSWM}}.

\subsection{Computational Complexity Analysis of \emph{CDP}}\label{de:Complexity_CDP}

\noindent\textsc{Proof:} First, for any agent $i$, the size of her feasible strategy set ($\mathcal{S}_{i}\left ( t \right)$) is $d_{i}^{in} + d_{i}^{out} + 1$. During each iteration, the modified utilities for all candidate strategies needs to be calculated for $n$ agents, and it takes time $\mathcal{O} \left [ \textstyle \sum_{i}  \left | \mathcal{S}_{i} \left ( t \right) \right | \right ] = \mathcal{O} \left (   2M + n \right ) = \mathcal{O} \left (   M \right )$, given that $M \gg n$ in the vast majority of real-world networks. Let $t^{Exp}$ denotes the expected number of iterations required for the \emph{best-response dynamics} to reach the termination condition, then the overall time complexity is $\mathcal{O} \left ( t^{Exp} M \right )$.

\subsection{Degenerates into the HK model}\label{de:degenerates_hk}

\noindent\textsc{Proof:} We first introduce a notion called \emph{Unit step function} $\varepsilon \left ( x \right ) = \left\{\begin{matrix}
    1, & if \; \; x > 0\\
    0, &  if \; \; x \le  0
\end{matrix}\right. $ Under the above conditions, $\Phi_{ij} \left ( T \right )$ and $\delta_{i}\left( T \right)$ can be written as

\begin{equation}
	\begin{aligned}
		\label{ao_2}
    \Phi_{ij}\left ( T \right ) = \frac{ \varepsilon \left \{ \mathbf{W}_{ij} \cdot \left [  \beta \gamma ^{T} -\left | \mathbf{x}\left ( T \right ) 1_{n}^{\mathsf{T}} - 1_{n}\mathbf{x}^{\mathsf{T}}\left ( T \right ) \right | \right ] \right \} }{\sum_{j\ne i}  \varepsilon \left \{ \mathbf{W}_{ij} \cdot \left [  \beta \gamma ^{T} -\left | \mathbf{x}\left ( T \right ) 1_{n}^{\mathsf{T}} - 1_{n}\mathbf{x}^{\mathsf{T}}\left ( T \right ) \right | \right ] \right \} + 1} 
    \\
    \delta_{i}\left ( T \right ) = \frac{ 1 }{\sum_{j\ne i} \varepsilon \left \{ \mathbf{W}_{ij} \cdot \left [  \beta \gamma ^{T} -\left | \mathbf{x}\left ( T \right ) 1_{n}^{\mathsf{T}} - 1_{n}\mathbf{x}^{\mathsf{T}}\left ( T \right ) \right | \right ] \right \} + 1 } 
   \end{aligned}
\end{equation}

Therefore, $\forall j\in \mathcal{N}_{i}^{T}$, we have $\Phi_{ij}\left ( T \right ) = \delta_{i}\left ( T \right ) = \frac{1}{\left | \mathcal{N}_{i}^{T} \right | + 1}$. And \emph{CAOFP} updates agent’s opinion using the following form:

\begin{equation}
	\begin{aligned}
		\label{ao_2}
    \mathbf{x}_{i} \left ( T+1 \right ) &= \delta_{i}\left ( T \right ) \mathbf{x}_{i}\left ( T \right ) + \sum_{j\ne i}\Phi_{ij}\left ( T \right )  \mathbf{x}_{j}\left ( T \right ) = \frac{1}{\left | \mathcal{N}_{i}^{\mathsf{T}} \right | + 1} \sum_{j \in \mathcal{N}_{i}^{\mathsf{T}} \cup \{i\} } \mathbf{x}_{j}\left ( T \right )
   \end{aligned}
\end{equation}

\subsection{Convergence of Opinion Formation Process}\label{de:Convergence_of_OFP}

\noindent\textsc{Proof:} Let $c = argmax_{a \in N_{i}^{\mathsf{T}}} \left | \mathbf{x}_{a}\left ( T \right ) - \mathbf{x}_{i}\left ( T \right )  \right | $, clearly $\left | \mathbf{x}_{c}\left ( T \right ) - \mathbf{x}_{i}\left ( T \right ) \right | < \beta \gamma ^{T}$. Since $\delta_{i} \left ( T \right ) \in \left ( 0,1 \right ]$ and ${\textstyle \sum_{j \ne i}}  \Phi _{ij} = 1-\delta _{i}$, it is apparent that
\begin{equation}
	\begin{aligned}
		\label{ao_2}
    \left | \mathbf{x}_{i}\left ( T+1 \right ) - \mathbf{x}_{i}\left ( T \right ) \right | 
    & = \left | \delta_{i} \left ( T \right ) \mathbf{x}_{i}\left ( T \right ) + \sum_{j\ne i}\Phi_{ij}\left ( T \right ) \cdot \mathbf{x}_{j}\left ( T \right ) - \mathbf{x}_{i}\left ( T \right ) \right | \\
    & =  \left | \sum_{j\ne i}\Phi_{ij}\left ( T \right ) \cdot \mathbf{x}_{j}\left ( T \right ) - \left ( 1- \delta_{i} \right ) \mathbf{x}_{i}\left ( T \right ) \right |  \\
    & < \left ( 1- \delta_{i} \right ) \left | \mathbf{x}_{c}\left ( T \right ) - \mathbf{x}_{i}\left ( T \right ) \right |   \\
    & \le \left ( 1- \delta_{i} \right )\beta \gamma ^{T}   < \beta \gamma ^{T}
   \end{aligned}
\end{equation}

\subsection{Convergence of Community merbership matrix}\label{de:Convergence_of_Community}

\noindent\textsc{Proof:} With Theorem \ref{theorem:5}, once the \emph{CAOFP} converges to the stable state, we have $ \left | \mathbf{x}_{i}\left (  T^{\ast } +1 \right ) - \mathbf{x}_{i}\left (  T^{\ast } \right ) \right | < \epsilon , \forall i \in N$. Set $\Delta _{i} \left ( T^{\ast } \right ) = \mathbf{x}_{i}\left (  T^{\ast } +1 \right ) - \mathbf{x}_{i}\left (  T^{\ast } \right ) $ and $\Delta _{j} \left ( T^{\ast } \right ) = \mathbf{x}_{j}\left (  T^{\ast } +1 \right ) - \mathbf{x}_{j}\left (  T^{\ast } \right ) $, clearly, $\Delta _{i} \left ( T^{\ast } \right ), \Delta _{j} \left ( T^{\ast } \right ) \in \left ( - \epsilon, \epsilon \right ) $.

\begin{equation}
	\begin{aligned}
		\label{ao_2}
    &\mathbf{Y}_{ij}\left ( T^{\ast } +1 \right ) - \mathbf{Y}_{ij}\left ( T^{\ast } \right ) \\
    &= \mathbf{W}_{ij} \cdot \left \{ \lambda_{i}\left [ \mathbf{D}_{ij}\left ( T^{\ast } +1 \right )- \mathbf{D}_{ij}\left ( T^{\ast } \right ) \right ] - ReLu\left [ \mathbf{D}_{ij}\left ( T^{\ast } + 1\right ) \right ] + ReLu\left [ \mathbf{D}_{ij}\left ( T^{\ast }\right ) \right ] \right \} \\
    &+ W_{ji} \cdot \left \{ \lambda_{j}\left [ \mathbf{D}_{ji}\left ( T^{\ast } +1 \right )- \mathbf{D}_{ji}\left ( T^{\ast } \right ) \right ] - ReLu\left [ \mathbf{D}_{ji}\left ( T^{\ast } + 1\right ) \right ] + ReLu\left [ \mathbf{D}_{ji}\left ( T^{\ast }\right ) \right ] \right \}
   \end{aligned}
\end{equation}

As $\mathbf{D}_{ij}\left ( T \right ) = \psi-\left|\mathbf{x}_{i}\left ( T \right )-\mathbf{x}_{j}\left ( T \right )\right|$, namely, 
\begin{equation}
	\begin{aligned}
		\label{D_T}
    \left | \mathbf{D}_{ij}\left ( T^{\ast } +1 \right ) - \mathbf{D}_{ij}\left ( T^{\ast } \right ) \right |
    &= \left | \left | \mathbf{x}_{i}\left (T^{\ast } \right) - \mathbf{x}_{j}\left (T^{\ast } \right) \right | - 
    \left | \mathbf{x}_{i}\left (T^{\ast } \right) - \mathbf{x}_{j}\left (T^{\ast } \right) + \Delta _{i} \left ( T^{\ast } \right ) - \Delta _{j} \left ( T^{\ast } \right )\right | \right | \\
    &\le \left | \left [ \mathbf{x}_{i}\left (T^{\ast } \right) - \mathbf{x}_{j}\left (T^{\ast } \right) \right ] - \left [ \mathbf{x}_{i}\left (T^{\ast } \right) - \mathbf{x}_{j}\left (T^{\ast } \right) \right ] - \left [ \Delta _{i} \left ( T^{\ast } \right ) - \Delta _{j} \left ( T^{\ast } \right ) \right ] \right |  \\
    &= \left | \Delta _{j} \left ( T^{\ast } \right )- \Delta _{i} \left ( T^{\ast } \right )  \right |
    \in \left [ 0, 2\epsilon \right )  \\
   \end{aligned}
\end{equation}

With Eq.~(\ref{D_T}), we can deduce that $-2\epsilon < \mathbf{D}_{ij}\left ( T^{\ast } +1 \right ) - \mathbf{D}_{ij}\left ( T^{\ast } \right ) < 2\epsilon$, further can be obtained that $ReLu\left[\mathbf{D}_{ij}\left ( T^{\ast }\right ) \right] - ReLu\left[\mathbf{D}_{ij}\left ( T^{\ast} + 1\right ) \right] < 2\epsilon$.

Because of $\mathbf{D}_{ji}\left ( T^{\ast } +1 \right ) = \mathbf{D}_{ij}\left ( T^{\ast } +1 \right )$ and $\mathbf{D}_{ji}\left ( T^{\ast } \right ) = \mathbf{D}_{ij}\left ( T^{\ast } \right )$, we obviously will have $-2\epsilon < \mathbf{D}_{ji}\left ( T^{\ast } +1 \right ) - \mathbf{D}_{ji}\left ( T^{\ast } \right ) < 2\epsilon$ and $ReLu\left[\mathbf{D}_{ji}\left ( T^{\ast }\right ) \right] - ReLu\left[\mathbf{D}_{ji}\left ( T^{\ast} + 1\right ) \right] < 2\epsilon$. Therefore, $\mathbf{Y}_{ij}\left ( T^{\ast } +1 \right ) - \mathbf{Y}_{ij}\left ( T^{\ast } \right ) < 2\epsilon \mathbf{W}_{ij}\left ( \lambda_{i} + 1\right ) + 2\epsilon \mathbf{W}_{ji}\left ( \lambda_{j} + 1\right ) = \mathcal{O}\left ( \epsilon  \right )$, i.e., matrix $\mathbf{Y} \left (  T \right ) $ will be steady when $T \ge T^{\ast } + 1$, such that $\mathbf{Y} \left ( T \right ) \approx \mathbf{Y} \left ( T +1 \right ) \approx \cdots$. Recall that $\mathbf{s} \left (  T^{\ast } + 1  \right ) $ is the pure Nash equilibrium w.r.t. \textbf{CFG}, we have

\begin{equation}
	\begin{aligned}
		\label{STABLE_COMMUNITY}
    u_{i} ^{\ast }  \left ( \mathbf{s} \left (  T^{\ast } + 1  \right ) \right ) 
    &= \sum_{j\ne i}\mathbf{s}_{i}\left (  T^{\ast } + 1  \right ) \mathbf{s}_{j}^{\mathsf{T}}\left (  T^{\ast } + 1  \right )\mathbf{Y}_{ij}\left (  T^{\ast } \right ) \\
    &\ge \sum_{j\ne i} e_{\left (  k \right )}^{\mathsf{T}} \mathbf{s}_{j}^{\mathsf{T}}\left (  T^{\ast } + 1  \right )\mathbf{Y}_{ij}\left (  T^{\ast } \right ), \forall e_{\left (  k \right )}^{\mathsf{T}} \in \mathcal{S}_{i}\left (  \tau  \right ), \forall i \in N
   \end{aligned}
\end{equation}

With Eq.~(\ref{STABLE_COMMUNITY}), one can easily deduce that the updated community membership matrix using will keep unchanged, such that $\mathbf{s} \left ( T \right ) = \mathbf{s} \left ( T +1 \right ) = \cdots$ when $T \ge T^{\ast } + 1$.

\subsection{Computational Complexity Analysis of \emph{GCAOFP}}\label{de:Complexity_GCAOFP}

\noindent\textsc{Proof:} During each iteration, the time cost of \textbf{\emph{Opinion Formation Process}} is $\mathcal{O} \left ( M \right )$. The time complexity of \emph{GCAOFP} during each period $t$ is the sum of the cost of the two phases: $\mathcal{O} \left ( t^{Exp} M \right ) + \mathcal{O} \left ( M \right ) = \mathcal{O} \left ( t^{Exp} M \right )$. Set $T^{Exp}$ is the expected number of iterations required by the convergence of \textbf{\emph{Opinion Formation Process}}, clearly $T^{Exp}$ is upper-bounded by $log_{\gamma }\frac{\epsilon }{\beta }$. Thus, the total time consumed by Algorithm \ref{alg} is $\mathcal{O} \left ( log_{\gamma } \frac{\epsilon }{\beta } t^{Exp} M \right ) = \mathcal{O} \left (t^{Exp} M \right )$.

\subsection{Evaluation Metrics.}\label{de:Evaluation_Metrics_NEW}

We attempt to define a social welfare gain indicator to measure the ability of \emph{GCAOFP} to improve social welfare with the given initial situation. According to Eq.~(\ref{utility_define}), at any time $T \in \mathbb{R}^{+}$, the overall social welfare $sw \left( T \right) = \sum_{i\in \mathcal{N} } u_{i}\left ( T \right ) = tr\left \{\mathbf{s}^{\mathsf{T}}\left ( T \right )\left [ \left ( \Lambda \mathbf{D}\left ( T \right ) - \mathbf{B}\left ( T \right ) \right ) \odot \mathbf{W} \right ] \mathbf{s}\left ( T \right ) \right \} + tr\left ( \mathbf{B}\left ( T \right ) \mathbf{W} \right ) $ has the maximum value $\sum_{i\in \mathcal{N} } \lambda _{i} \psi \sum_{j\ne i} \textbf{W}_{ij} = \lambda \psi W$, at which time all agents belong to the same community and have the same opinion, which corresponds to the consensus state eventually reached by models such as \emph{DeGroot}. Meanwhile, $sw \left( T \right)$ will also has the minimum value $ \left ( \psi - 1 \right ) \lambda W$, which means that all agents belong to the same community, but all agents hold extreme opinions, i.e., $\textbf{x}^{\star } = \left \{ 0, 1 \right \}^{n}$ and any neighboring agents have opposite opinions. Therefore, the maximum gain in social welfare is $\lambda \psi W - \left ( \psi - 1 \right ) \lambda W = \lambda W$, and we thus define the \textbf{O}verall \textbf{S}ocial \textbf{W}elfare \textbf{G}ain (\emph{OSWG}) as

\begin{equation}
	\begin{aligned}
		\label{OSWG}
    OSWG = \frac{sw\left ( T^{\ast} \right ) - sw\left ( 0 \right )}{\lambda W},
   \end{aligned}
\end{equation}

\noindent where $sw\left ( 0 \right )$ is calculated from the given opinion vector and community structure, and $sw\left ( T^{\ast} \right )$ can be calculated using the opinion vector and community structure output from \emph{GCAOFP}. We also defined \textbf{R}elative \textbf{C}onsensus \textbf{S}ocial \textbf{W}elfare (\emph{RCSW}) to reflect the level of consensus reached among agents at the eventual steady state of \emph{GCAOFP}:

\begin{equation}
	\begin{aligned}
		\label{OSWG}
    RCSW = \frac{sw\left ( T^{\ast} \right )}{\lambda \psi W}  
   \end{aligned}
\end{equation}

Similarly, we defined \textbf{I}nitial \textbf{C}onsensus \textbf{S}ocial \textbf{W}elfare (\emph{ICSW}) to reflect the level of consensus reached among agents at the initial state:

\begin{equation}
	\begin{aligned}
		\label{OSWG}
    ICSW = \frac{sw\left ( 0 \right )}{\lambda \psi W}  
   \end{aligned}
\end{equation}

Finally, We employ evaluation metrics \emph{Average Consensus Level} (\emph{ACL}) to quantitatively measure the level of consensus among members within the identified communities (for standard opinion dynamics models, we treat the entire network as a community), 

\begin{equation}
	\begin{aligned}
		\label{ACL_DEFINE}
    ACL = \frac{1}{k}\sum_{k} CL_{k}= \frac{1}{k}\sum_{k}\sum_{i\in \mathcal{C}_{k}} \frac{1- \left | \mathbf{x}_{i}^{\star} - \mathbf{c}_{k}^{\star} \right |}{\left | \mathcal{C}_{k} \right |},
   \end{aligned}
\end{equation}

\noindent where $\mathcal{C}_{k}$ indicates the $k$-th community in the equilibrium state and $\mathbf{c}_{k}^{\star} = \frac{1}{\left|\mathcal{C}_{k} \right|}\sum_{i\in \mathcal{C}_{k}} \mathbf{x}_{i}^{\star}$ denotes the average opinion of all members within $\mathcal{C}_{k}$. Obviously, the closer the \emph{ACL} value is to 1, the more similar the opinions of agents within the same community.

\subsection{The effect of network density on robustness}\label{de:density_robustness}
\begin{figure}[htbp]
\center
\linespread{0.1}
\includegraphics[width=1\textwidth]{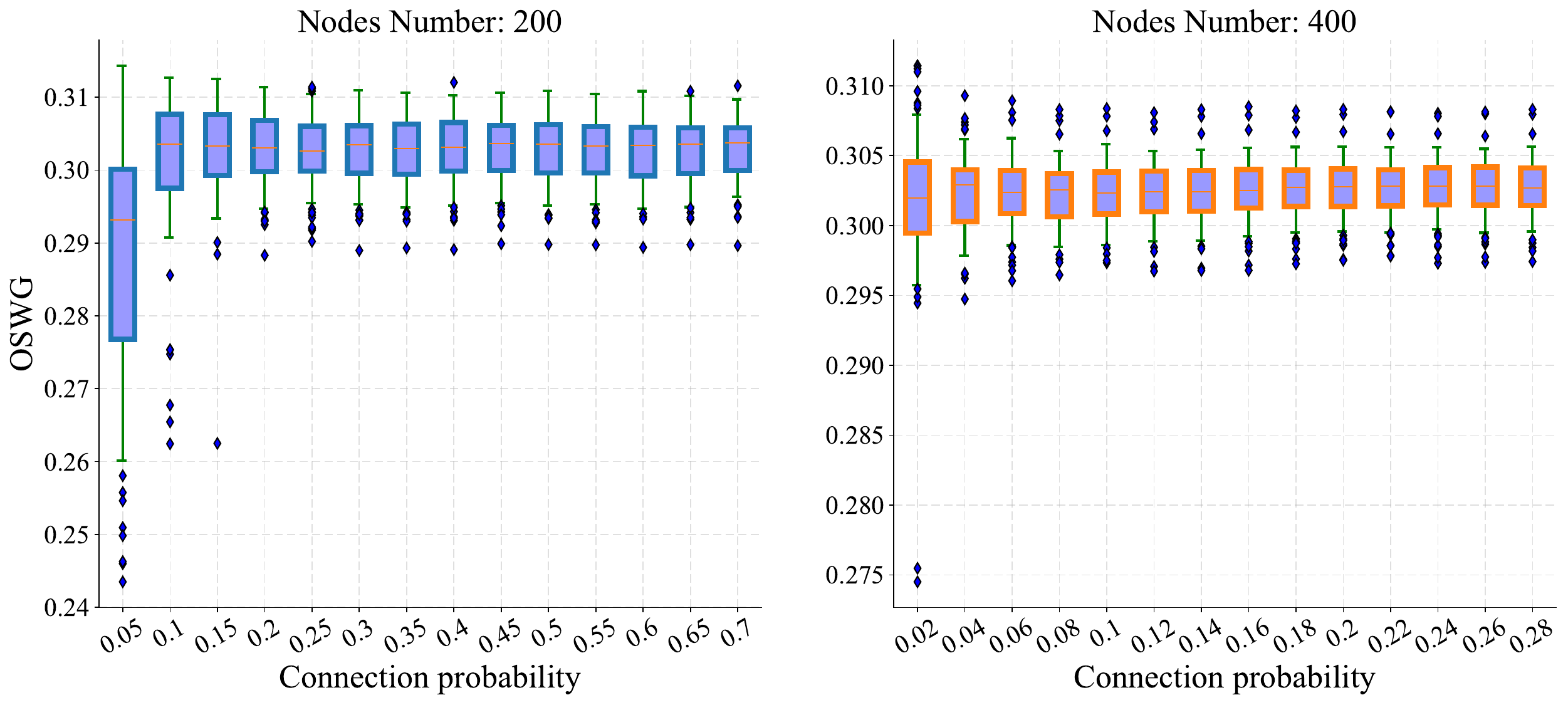}
\caption{The boxplots of \emph{OSWG} (100 implements) of \emph{GCAOFP} on networks with different densities.}
\label{fig:density_robustness}
\end{figure}

We study the effect of network density by controlling the connection probability of \emph{ER} networks. Specifically, we configure the same initial opinions and community member matrices for networks with the same number of agents, set the same parameters ($\lambda=1.4$ and $\psi=0.4$), and then continuously increase the connection probability between nodes and observe the changes in \emph{OSWG}. Fig.\ref{fig:density_robustness} summarizes the experimental results, and it is clear that the variance of \emph{OSWG} decreases as the network density increases, indicating that \emph{GCAOFP} is increasingly robust to the initial state.

\subsection{Initial Consensus Social Welfare}\label{de:Initial_consensus}

Fig.\ref{consensus_draw} shows the low-quality community partitions obtained by \emph{GK-Means} and \emph{LPA-HK} in order to achieve convergence of agents' opinions. We can intuitively see: \circled{1} the opinions of agents within the same community in (a) are basically the same, but the opinions of agents belong to different communities are obviously different, indicating that the opinions have achieved local convergence; \circled{2} the opinions of agents in different communities in (b) and (c) are not much different, indicating that the opinions have reached a consensus; \circled{3} all the three community division results are of low quality and are very different from the ground-truth communities.


\subsection{Initial State}\label{de:Initial_state}

\begin{figure}[htbp]
	\centering

    \begin{minipage}{0.99\linewidth}
		\centering
		\subfloat[\emph{Karate} divided by \emph{GK-Means}]{
        \includegraphics[width=2.5in]{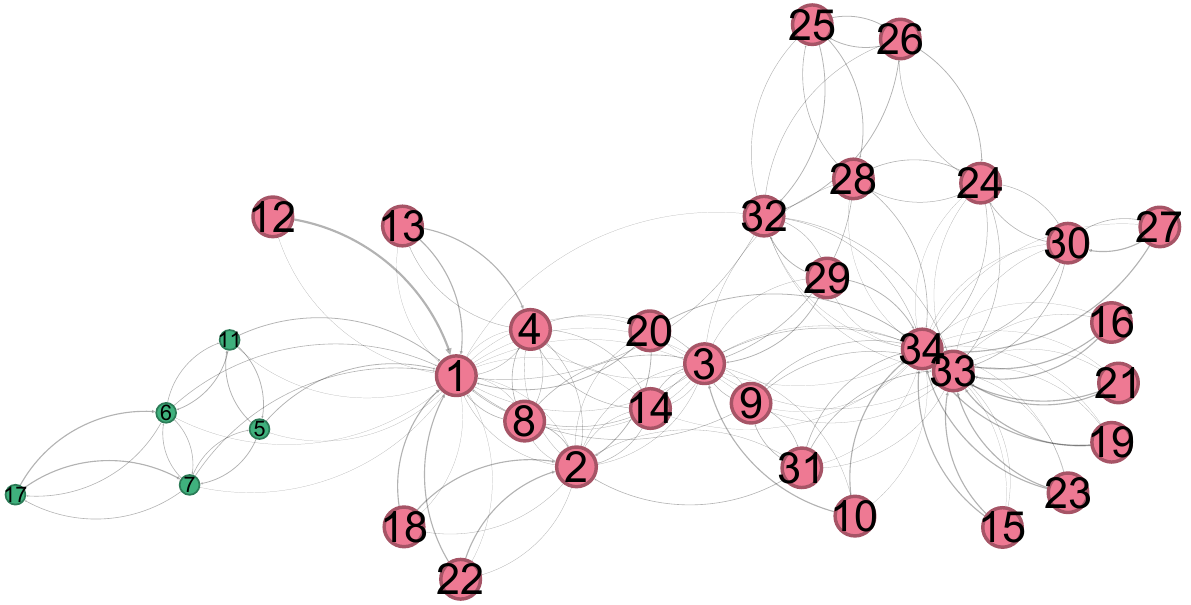}
        }
        \subfloat[\emph{Karate} divided by \emph{LPA-HK}]{
	   \includegraphics[width=2.5in]{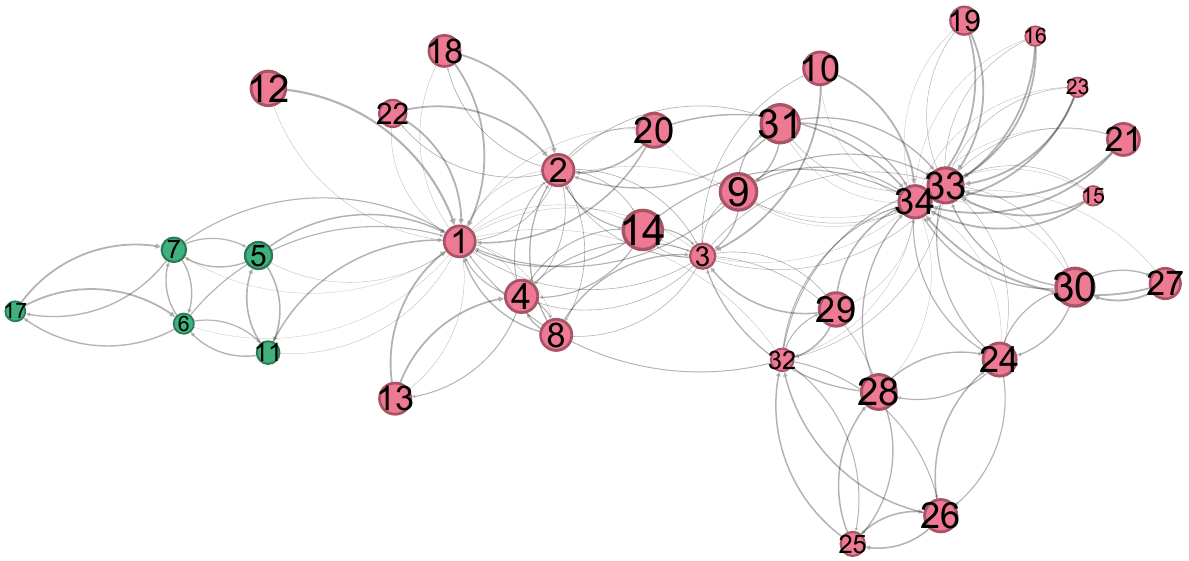}
        }
        \quad
        \subfloat[\emph{Polbooks} divided by \emph{LPA-HK}]{
	   \includegraphics[width=2.5in]{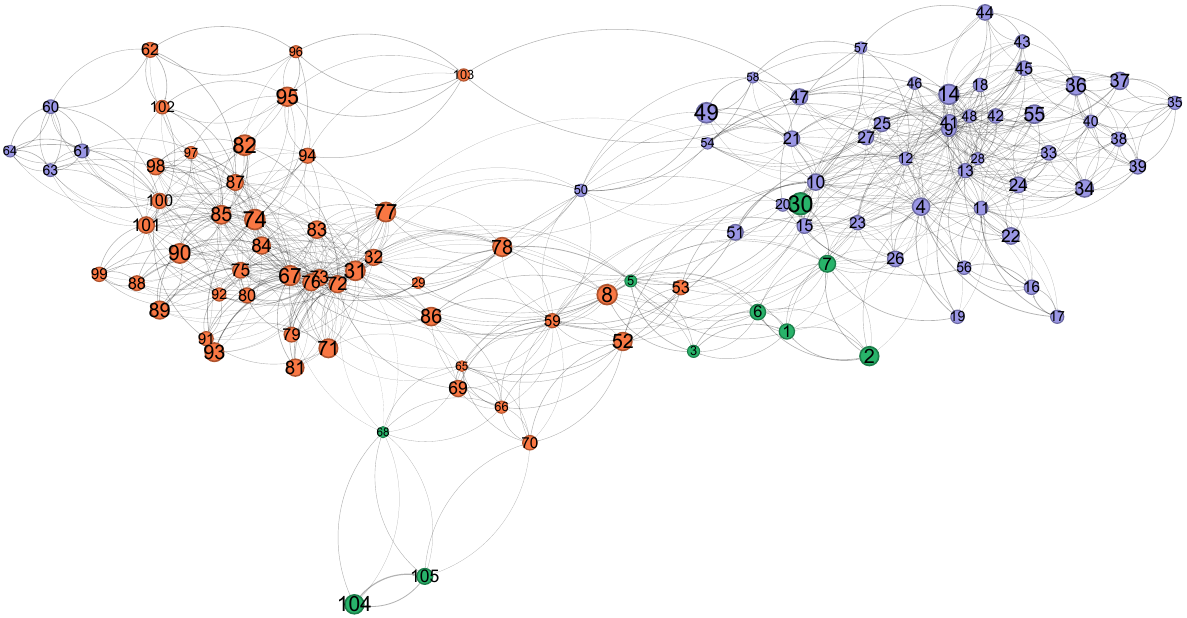}
        }
        \caption{The communities detected by \emph{GK-Means} and \emph{LPA-HK}. The size of nodes reflect their final opinions, the color of nodes are determined by community labels, and the link thickness between nodes indicate the impact of the target node on the source node.}
        \label{consensus_draw}
	\end{minipage}

    \begin{minipage}{0.99\linewidth}

    \centering
    \linespread{0.1}
    \includegraphics[width=1\textwidth]{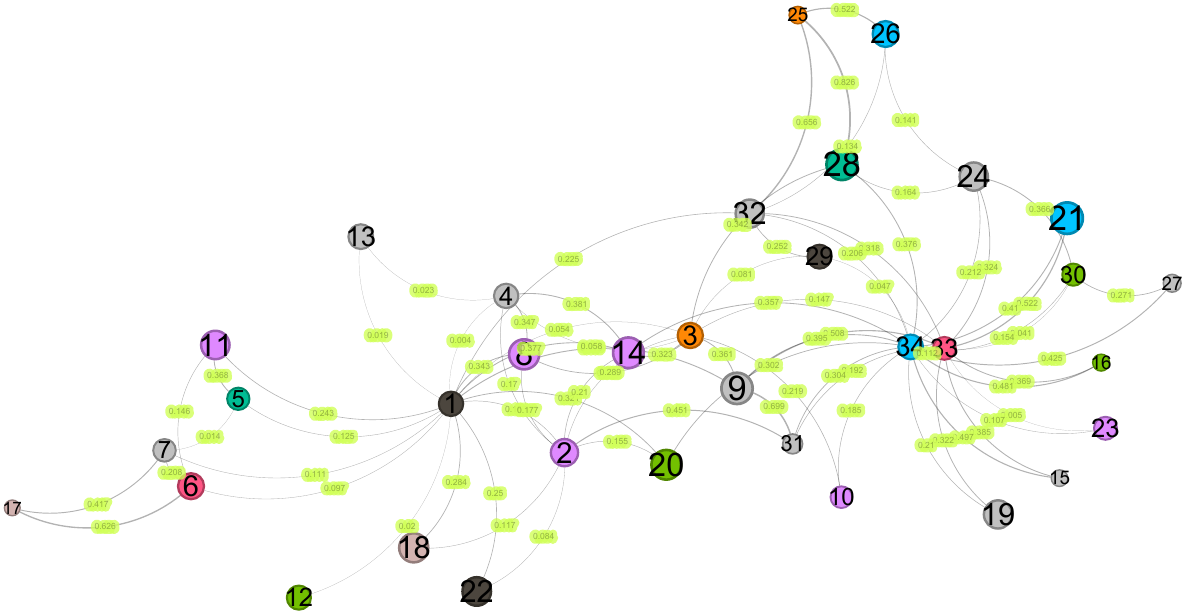}
    \caption{The initial state of the \emph{Karate} network. The size of nodes reflect their initial opinions, the color of nodes are determined by initial community labels, the link thickness between nodes indicate the absolute value of the difference in opinions and we have also clearly marked this difference in opinion with numbers.}
    \label{fig:ini_karate_opinion}
    
    \end{minipage}

\end{figure}

Fig.\ref{fig:ini_karate_opinion} visualizes the initial state of the \emph{Karate} network, and it is clear that the initial community division was chaotic, with many agents being divided into different communities from their neighbors. In addition, initial opinions vary greatly between neighboring agents, the most typical examples include agents 17 and 25.


\end{appendices}

\clearpage


\bibliography{sn-bibliography}

\end{document}